\documentclass[10pt,letterpaper,twocolumn]{article}

\usepackage[utf8]{inputenc}
\usepackage{amsmath}
\usepackage{booktabs}
\usepackage{amsfonts}
\usepackage{graphicx}
\usepackage{stfloats}
\usepackage[hidelinks]{hyperref}
\usepackage{multirow}
\usepackage{xcolor}
\usepackage{url}
\usepackage[nameinlink,capitalize,noabbrev]{cleveref}
\usepackage[font=small,labelfont=bf,labelsep=period]{caption}
\usepackage[letterpaper,portrait,bottom=1.0in,top=0.6in,left=0.7in,right=0.7in,footskip=0.5in]{geometry}
\usepackage[round]{natbib}

\hypersetup{colorlinks=true, linkcolor=black, urlcolor=blue, citecolor=black}
\DeclareMathOperator*{\argmin}{arg\,min}
\newcommand{\subpara}[1]{\vspace{2mm}\noindent\textbf{#1:}\hspace{0mm}}

\renewenvironment{abstract}
    {
    \small
    {\fontsize{11pt}{12pt}\selectfont\textbf{\abstractname}}\vspace{5pt}
    \list{}{
    \setlength{\leftmargin}{0mm}
    \setlength{\rightmargin}{0mm}
    }
    \item\relax
    }
    {\endlist}

\makeatletter
\renewcommand{\maketitle}{
\bgroup\setlength{\parindent}{0pt}
\begin{flushleft}
\vspace{20pt}
{\fontsize{20pt}{12pt}\selectfont\textbf{\@title}}
\\\vspace{20pt}
\@author
\\\vspace{12pt}
\@date
\\\vspace{12pt}
\end{flushleft}\egroup
}
\makeatother

\title{SynthStrip: Skull-Stripping for Any Brain Image}

\author{
    \textbf{Andrew~Hoopes}~$^{1}$,
    \textbf{Jocelyn~S.~Mora}~$^{1}$,
    \textbf{Adrian~V.~Dalca}~$^{1-3}$,
    \textbf{Bruce~Fischl}~$^{1-4,\,\dagger}$,
    \textbf{Malte~Hoffmann}~$^{1,2,\,\dagger}$
}

\date{
    \footnotesize
    $^1$ Athinoula A.\ Martinos Center for Biomedical Imaging, Massachusetts General Hospital, Charlestown, MA, USA\\
    $^2$ Department of Radiology, Harvard Medical School, Boston, MA, USA\\
    $^3$ Computer Science and Artificial Intelligence Lab, Massachusetts Institute of Technology, Cambridge, MA, USA\\
    $^4$ Harvard-MIT Division of Health Sciences and Technology, Cambridge, MA, USA\\
    \vspace{0.2cm}
    $^{\dagger}$ These authors share senior authorship.
}

\begin{document}

\twocolumn[
\begin{@twocolumnfalse}
\maketitle

\noindent\rule{\textwidth}{0.5pt}
\vspace{0mm}

\begin{abstract}
\textbf{The removal of non-brain signal from magnetic resonance imaging (MRI) data, known as skull-stripping, is an integral component of many neuroimage analysis streams. Despite their abundance, popular classical skull-stripping methods are usually tailored to images with specific acquisition properties, namely near-isotropic resolution and T1-weighted (T1w) MRI contrast, which are prevalent in research settings. As a result, existing tools tend to adapt poorly to other image types, such as stacks of thick slices acquired with fast spin-echo (FSE) MRI that are common in the clinic. While learning-based approaches for brain extraction have gained traction in recent years, these methods face a similar burden, as they are only effective for image types seen during the training procedure. To achieve robust skull-stripping across a landscape of imaging protocols, we introduce SynthStrip, a rapid, learning-based brain-extraction tool. By leveraging anatomical segmentations to generate an entirely synthetic training dataset with anatomies, intensity distributions, and artifacts that far exceed the realistic range of medical images, SynthStrip learns to successfully generalize to a variety of real acquired brain images, removing the need for training data with target contrasts. We demonstrate the efficacy of SynthStrip for a diverse set of image acquisitions and resolutions across subject populations, ranging from newborn to adult. We show substantial improvements in accuracy over popular skull-stripping baselines -- all with a single trained model. Our method and labeled evaluation data are available at~\url{https://w3id.org/synthstrip}.
}

\end{abstract}

\vspace{1pt}
\noindent\rule{\textwidth}{0.5pt}
\vspace{0pt}

\section{Introduction}

\end{@twocolumnfalse}
]

\noindent
Skull-stripping, also known as brain extraction, involves the removal of non-brain tissue signal from magnetic resonance imaging (MRI) data. This process is useful for anonymizing brain scans and a fundamental component of many neuroimage analysis pipelines, such as FreeSurfer~\citep{fischl2012}, FSL~\citep{jenkinson2012fsl}, AFNI~\citep{cox1996afni}, and ANTs~\citep{avants2011reproducible}. These packages include tools that typically require brain-extracted input images and might perform inaccurately, or even fail, without removal of irrelevant and distracting tissue. One such class of algorithms that benefits from this systematic tissue extraction is image registration, a core element of atlas-based segmentation and other analyses. Nonlinear registration~\citep{ashburner2007fast,avants2008symmetric,rueckert1999nonrigid,vercauteren2009diffeomorphic} estimates local deformations between pairs of images, and these algorithms tend to produce more accurate estimates when they can focus entirely on the anatomy of interest~\citep{klein2009evaluation,ou2014comparative}. Similarly, skull-stripping increases the reliability of linear registration~\citep{cox1999real,friston1995spatial,hoffmann2015survey,jenkinson2001global,jiang1995motion,modat2014global,reuter2010} by excluding anatomy that deforms non-rigidly, such as the eyes, jaw, and tongue~\citep{andrade2018practical,fein2006statistical,fischmeister2013benefits,hoffmann2020real}.

Classical skull-stripping techniques are well-explored and widespread, but popular methods are often tailored to images with specific modalities or acquisition properties. Most commonly, these methods focus on three-dimensional (3D) T1-weighted (T1w) MRI scans acquired with MPRAGE sequences~\citep{marques2010mp2rage,mugler1990,van2008brain}, which are ubiquitous in neuroimaging research. While some skull-stripping tools accommodate additional contrasts, these methods are ultimately limited to a predefined set of viable image types and do not properly adapt to inputs outside this set. For example, skull-stripping tools developed for near-isotropic, adult brain images may perform poorly when applied to infant subjects or clinical scans with thick slices, such as stacks of 2D fast spin-echo (FSE) acquisitions.

When a suitable brain extraction method is not available for a particular scan type, a common workaround involves skull-stripping a compatible image of the same subject and computing a co-registration to propagate the extracted brain mask to the target image of interest~\citep{iglesias2011robust}. Unfortunately, an accurate intra-subject alignment can require significant manual tuning because the target image still includes extra-cerebral matter that may impede linear registration quality~\citep{reuter2010}. Crucially, this procedure also requires the existence of an additional, strip-able image, often a high-resolution isotropic T1w or T2-weighted (T2w) scan, which is rare, for example, in clinical screening protocols, introducing a barrier to the clinical adoption of analysis tools.

\begin{figure*}[t]
    \centering
    \includegraphics[width=\textwidth]{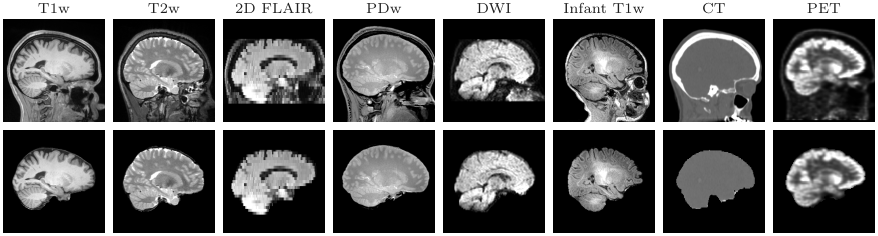}
    \caption{Examples of SynthStrip brain extractions (bottom) for a wide range of image acquisitions and modalities (top). Powered by a strategy for synthesizing diverse training data, SynthStrip learns to skull-strip brain images of any type.}
    \label{fig:example_skullstrips}
\end{figure*}

While classical algorithms for skull-stripping are limited by their assumptions about the spatial features and intensity distributions in the input images, supervised deep-learning approaches, which leverage convolutional neural networks (CNNs), can, in principle, learn to extract a region of interest from any image type given sufficient anatomical contrast and resolution. In practice, these networks achieve high accuracy for data types observed during training, but their performance often deteriorates on images with characteristics unseen during training~\citep{hendrycks2021iccv,hoffmann2021learning,jog2019psacnn,karani2018lifelong}. In consequence, robust, supervised learning-based approaches depend on the availability of a representative training dataset that contains accurate ground-truth annotations and exposes the network to a landscape of image types. While numerous public datasets provide access to widely used MRI acquisitions for which target brain masks can be easily derived with classical methods, curating a diverse training dataset with uncommon sequences and sufficient anatomical variability is a challenging task that requires substantial human effort. As a result, current deep-learning skull-stripping methods are trained with few different data types and deliver state-of-the-art results only for particular subsets of image characteristics~\citep{hwang2019,kleesiek2016deep,salehi2017auto}.

Recently, a novel learning strategy alleviates the requirement for representative acquired training data by optimizing networks with a wide array of synthetic images, each generated directly from a precomputed label map~\citep{billot2020learning,hoffmann2021learning}. This synthesis scheme enables networks to accurately carry out tasks on any image type at evaluation-time without ever sampling real target acquisitions during training, and it has been effectively employed for segmentation~\citep{billot2020learning} and deformable image registration~\citep{hoffmann2021learning}. To build on deep-learning methods for brain extraction while addressing their shortcomings, we adapt the synthesis technique and introduce SynthStrip, a flexible brain-extraction tool that can be deployed universally on a variety of brain images. By exposing a CNN to an arbitrary and deliberately unrealistic range of anatomies, contrasts, and artifacts, we obtain a model that is agnostic to acquisition specifics, as it never samples any real data during training. Consequently, this scheme enables SynthStrip to extract the brain from a wide array of neuroimaging data types, and we demonstrate its viability and improvement over popular baselines using a varied test set that spans both research scans and clinical exams (Figure~\ref{fig:example_skullstrips}). The test set includes T1w, T2w, T2w fluid attenuated inversion recovery (T2-FLAIR), and proton-density (PDw) contrasts as well as clinical FSE scans with slices and high in-plane resolution, and low-resolution EPI, ranging across age and pathology. We demonstrate the ability of SynthStrip to generalize beyond structural MRI, to MR angiography (MRA), diffusion-weighted imaging (DWI), fluorodeoxyglucose positron emission tomography (FDG-PET), and even computed tomography (CT). We make our validation set publicly available to promote further development and evaluation of brain-extraction tools.

\vspace{10pt}
\section{Related Work}

In this section, we briefly review the automated brain-extraction techniques that we use as baseline methods. We include both classical and deep-learning baselines introduced over the last two decades, focusing in particular on those with high efficacy and popularity in the research domain. For an exhaustive overview of skull-stripping methods, see \cite{fatima2020state}.

\subsection{Classical Skull-Stripping}

Classical, or traditional, algorithms that remove non-brain image signal vary substantially in their implementation~\citep{cox1996afni,eskildsen2012beast,iglesias2011robust,roy2017,segonne2004hybrid,shattuck2001,smith2002fast}. One common class of approaches leverages a deformable mesh model to reconstruct a smooth boundary of the brain matter surface. The widely-used Brain Extraction Tool (BET; \citealp{smith2002fast}), distributed as part of the FSL package~\citep{jenkinson2012fsl}, utilizes this technique by initializing a spherical mesh at the barycenter of the brain and projecting mesh vertices outwards to model the brain border. Since BET uses locally adaptive intensity thresholds to distinguish brain and non-brain voxels, it generalizes to a variety of contrasts, such as T1w, T2w, and PDw. To prevent surface leaks beyond the brain boundary, 3dSkullStrip, a component of AFNI~\citep{cox1996afni}, extends the BET strategy by considering information on the surface exterior, accounting for eyes, ventricles, and skull.

The popular hybrid approach~\citep{segonne2004hybrid} available in FreeSurfer also leverages a deformable surface paradigm, combing it with a watershed algorithm and statistical atlas to improve robustness. First, the watershed establishes an estimate of the white-matter mask, which is then refined to the brain boundary using a surface mesh expansion. A probabilistic atlas of intensity distributions helps prevent outliers during mesh fitting, and erroneous brain mask voxels are removed during post-processing via a graph cuts algorithm~\citep{greig1989exact,sadananthan2010skull} that thresholds the cerebrospinal fluid (CSF). While effective, this technique is optimized only for images with T1w contrast, since it relies on the underlying assumption that white matter is surrounded by darker gray matter and CSF. Another hybrid approach, ROBEX~\citep{iglesias2011robust}, exploits a joint generative-discriminative model. A Random Forest classification~\citep{breiman2001random} detects the brain contour, which is used to fit a point-distribution model to the brain target. The skull-stripping tool BEaST~\citep{eskildsen2012beast} builds on patch-based, non-local segmentation techniques~\citep{coupe2010nonlocal,coupe2011patch,roy2017} and assigns a label to each voxel by comparing its local neighborhood to patches in a reference set with prior labels. With the exception of BET and 3dSkullStrip, all of these tools were specifically developed for T1w images.

\subsection{Deep-Learning Approaches}

Innovations in deep-learning have gained popularity as methodological building blocks for an array of tasks in medical image analysis, including skull-stripping. Various learning-based extraction methods have been proposed, demonstrating accuracy and speed that often out-perform their classical counterparts. These models are optimized in a supervised fashion, using a set of acquired training images with corresponding ground-truth brain masks, derived through classical methods or manual segmentation. An early, cross-contrast approach, Deep MRI Brain Extraction (DMBE)~\citep{kleesiek2016deep}, trains a 3D CNN on combinations of T1w, T2w, and FLAIR contrasts and matches the accuracy of classical baselines for several datasets, including clinical scans with brain tumors. Conversely, Auto-Net~\citep{salehi2017auto} introduces two separate 2.5D architectures that skull-strip volumes by individually segmenting sagittal, coronal, and transverse views of same image and fusing the predictions with an auto-context algorithm~\citep{tu2009auto}. The first architecture leverages convolutions on single-resolution voxel-wise patches, while the second utilizes a scale-space U-Net architecture~\citep{ronneberger2015} to predict the brain mask. Auto-Net is effective for both adult and neonatal brain scans but only trained with T1w images. CONSNet~\citep{lucena2019} similarly leverages a 2D U-Net, applied across image slices in each plane, to strip 3D T1w images. More recently, implementations using full 3D U-Nets~\citep{hsu2020,hwang2019} have robustly matched or exceeded start-of-the-art brain-extraction performance.

\subsection{Contribution}

SynthStrip builds on a solid foundation laid by prior studies of deep-learning algorithms for brain extraction, enabling us to choose among network architectures well suited for this particular task. We emphasize that our goal is not to compare or make claims on the optimality of specific architectures -- the discussed algorithms may perform equally well. Instead, our focus is on exploiting a novel training strategy using synthetic data only, to build an easy-to-use skull-stripping tool that alleviates the requirement of expanding the training set and re-optimizing network weights every time a new image type is to be supported.

\section{Method}

To predict robust brain masks for an array of real image types, we train a deep convolutional neural network on a vast landscape of images synthesized with a deliberately unrealistic range of anatomies, acquisition parameters, and artifacts. From a dataset~$\mathcal{D}$ of precomputed, whole-head segmentations with brain and non-brain tissue labels, we sample a segmentation~$s \in \mathcal{D}$ at each optimization step and use it to generate a gray-scale head scan~$x$ with randomized acquisition characteristics. In effect, this paradigm synthesizes a stream of training images used to optimize a SynthStrip network~$g_\theta$, with trainable parameters~$\theta$, in a supervised fashion:
\begin{equation}
    \hat{\theta} = \argmin_{\theta} \Big [ ~ \mathbb{E}_{\mathcal{D}} ~ [ ~ \mathcal{L}(y, \hat{y}) ~ ] ~ \Big ],
\end{equation}
where~$y$ is the predicted brain mask, ~$\hat{y}$ is the target brain mask derived by merging the brain labels of~$s$, and~$\mathcal{L}$ is the loss function that measures similarity between~$y$ and~$\hat{y}$.

\begin{figure}[t]
    \centering
    \includegraphics[width=\columnwidth]{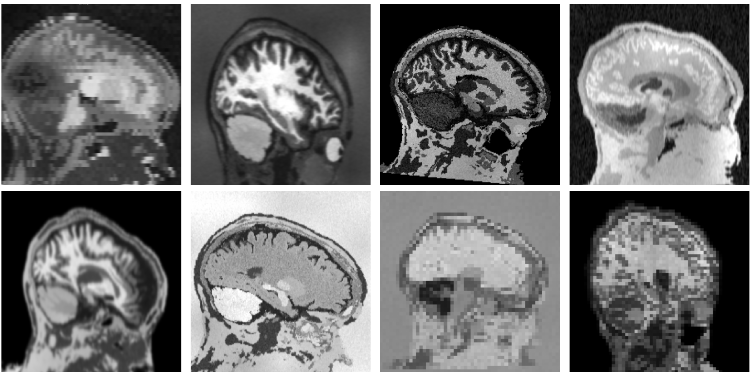}
    \caption{Samples of synthetic images used for SynthStrip training. To encourage the network to generalize, we synthesize images that far exceed the realistic range of whole-brain acquisitions. In this figure, each brain image is generated from the same label map. In practice, we use label maps from several different subjects.}
    \label{fig:synthesis_examples}
\end{figure}

\subsection{Synthesis}

Building from previous work~\citep{billot2020learning,hoffmann2021learning}, we use a generative model to synthesize a stream of random images with substantial anatomical and intensity variation, as exhibited in Figure~\ref{fig:synthesis_examples}. At each training step, parameters that dictate synthesis components are randomly sampled from predetermined ranges and probability distributions explicitly defined in Table~\ref{table_hyperparams}. We emphasize that while the generated scans can appear implausible, these training images do not need to be realistic in order for the SynthStrip model to accurately generalize to real images at test-time.

To generate a gray-scale image~$x$ from a whole-head anatomical segmentation~$s$, we first create spatial variability to subject the network to a landscape of possible head positions and anatomical irregularities. This is accomplished by manipulating~$s$ with a spatial transformation~$t$, composed of an affine transform (with random translation, scaling, and rotation) and a nonlinear deformation. The deformation is generated by sampling random 3D displacement vectors from a normal distribution, with random scale, at an arbitrarily low image resolution. This random displacement field is vector-integrated, using five~\textit{scaling and squaring} steps to encourage a diffeomorphic warp~\citep{arsigny2006log,dalca2019varreg}, and tri-linearly resampled to match the resolution of~$s$. After applying the randomized transform, the resulting segmentation~$s_t$ serves as the basis for deriving the image~$x$ and target brain mask~$\hat{y}$, which is obtained by merging the labels of~$s_t$ into brain and non-brain classes.

\begin{table}[]
\centering
\small
\begin{tabular}{ll}
\toprule
\textbf{Synthesis hyperparameter} & \textbf{Sampling range} \\
\toprule
Affine translation & 0--50 \textit{mm} \\
\hline \\ [-2ex]
Affine rotation & 0--45$^\circ$ \\
\hline \\ [-2ex]
Affine scaling & 80--120\% \\
\hline \\ [-2ex]
Deformation voxel length & 8--16 \textit{mm} \\
\hline \\ [-2ex]
Deformation SD & 0--3 \textit{mm} \\
\hline \\ [-2ex]
Label intensity mean & 0--1 \\
\hline \\ [-2ex]
Label intensity SD & 0--0.1 \\
\hline \\ [-2ex]
Bias field voxel length & 4--64 \textit{mm} \\
\hline \\ [-2ex]
Bias field SD & 0--0.5 \\
\hline \\ [-2ex]
Exponentiation parameter~$\gamma$ & -0.25--0.25 \\
\hline \\ [-2ex]
FOV cropping (any axis) & 0--50 \textit{mm} \\
\hline \\ [-2ex]
Down-sample factor~$r$ (any axis) & 1--5 \\
\bottomrule
\end{tabular}
\caption{Uniform hyperparameter sampling ranges used for synthesizing a training image from a source segmentation map. The specific values were chosen by visual inspection of the generated images to produce a landscape of image contrasts, anatomies, and acquisition characteristics that far exceed the realistic range of medical images. We sample fields with isotropic voxels of the indicated side length, where SD abbreviates standard deviation.}
\label{table_hyperparams}
\end{table}

To compute~$x$, we consider a Bayesian model of MR contrast, which assumes that the voxel intensity of each tissue type in the image can be represented by a single Gaussian distribution. Reversing this generalization, we assign a random distribution of tissue intensity to every anatomical label in~$s_t$ and use this artificial mixture model to attain an image with arbitrary contrast by replacing each label voxel in~$s_t$ with a random value drawn from its corresponding intensity distribution. Following the synthesis, we aim to simulate various artifacts and geometric properties that might exist across modality and acquisition type. First, we corrupt the image with a spatially varying intensity bias field, generated by resizing a low-resolution image sampled from a normal distribution with zero mean. The corrupted image is computed by an element-wise multiplication with the voxel-wise exponential of the bias field. Second, we perform gamma augmentation by globally exponentiating all voxels with a single value~$\exp(\gamma)$, where~$\gamma$ is a normally sampled parameter. Lastly, to account for scans with a partial field of view (FOV) and varied resolution, we randomly crop the image content and down-sample along an indiscriminate set of axes. Before down-sampling by an arbitrary factor~$r$, we simulate partial-volume effects by blurring the image using a Gaussian kernel with standard deviation~$\sigma = r / 4$. The image cropping and down-sampling components are applied with a 50\% probability rate during synthesis.

\begin{figure*}[t]
    \centering
    \includegraphics[width=\textwidth]{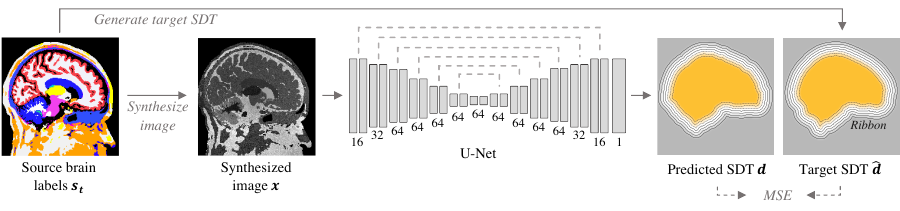}
    \caption{SynthStrip training framework. At every optimization step, we sample a randomly transformed brain segmentation $s_t$, from which we synthesize a gray-scale image~$x$ with arbitrary contrast. The skull-stripping 3D U-Net receives~$x$ as input and predicts a thresholded signed distance transform (SDT) $d$ representing the distance of each voxel to the skull boundary. The U-Net consists of skip-connected, multi-resolution convolutional layers illustrated by gray bars, with their number of output filters indicated below. We train SynthStrip in a supervised fashion, maximizing the similarity between~$d$ and the ground-truth SDT~$\hat{d}$ within a ribbon of set distance around the brain and derived directly from the segmentation labels of~$s_t$.}
    \label{fig:training_schematic}
\end{figure*}

\subsection{Loss}

We optimize~$g_\theta$ using a loss function~$\mathcal{L}$ that measures the similarity between predicted and target brain masks. Unless otherwise stated, we employ a loss~$\mathcal{L} = \mathcal{L}_{sdt}$ that encourages the network to predict a signed distance transform (SDT)~$d$ representing the minimum distance (in \textit{mm}) to the skull boundary at each voxel. Distances are positive within the brain and negative outside, facilitating the extraction of a binary brain mask~$y$ from~$d$ at test-time by simple thresholding. The training paradigm is outlined in Figure~\ref{fig:training_schematic}. During training, an exact target Euclidean SDT~$\hat{d}$ is computed from the target brain mask~$\hat{y}$, and the similarity between~$d$ and~$\hat{d}$ is measured by their mean squared difference (MSE). To concentrate optimization gradients to pertinent regions of the image during training,~$\hat{d}$ is banded such that voxel distances~$\hat{d}_i$ do not surpass a discrete threshold~$h$, and all voxels that exceed the distance~$h$ are down-weighted in the MSE computation by a factor~$b$. Therefore,
\begin{equation}
    \mathcal{L}_{sdt} = \frac{\sum_{i \in \mathcal{P}}{w_i (d_i - \hat{d}_i) ^ 2}}{\sum_{i \in \mathcal{P}}{w_i}}, \hspace{10pt} w_i =
    \begin{cases}
        b & \text{if~$|\hat{d}_i| > h$,} \\
        1 & \text{otherwise,}
    \end{cases}
\end{equation}
where~$i$ represents a voxel in the spatial image domain~$\mathcal{P}$, $h = 5$~\textit{mm}, and~$b = 0.1$ in our experiments, optimally determined via a grid search.

As a complimentary analysis, we compare the distance-based loss~$\mathcal{L}_{sdt}$ against a soft Dice loss~\citep{dice1945,milletari2016v}, which is commonly used to optimize image segmentation models and quantifies volume overlap for pairs of labels. We define the loss~$\mathcal{L}_{dice}$ as
\begin{equation}
    \mathcal{L}_{dice} = \frac{| ~ y^j \odot \hat{y}^j ~ |}{| ~ y^j \oplus \hat{y}^j ~ |} + \
    \frac{| ~  y^k \odot \hat{y}^k ~ |}{| ~ y^k \oplus \hat{y}^k ~ |},
\end{equation}
where~$y^j$ and~$\hat{y}^j$ represent brain label maps,~$y^k$ and~$\hat{y}^k$ represent non-brain label maps, and~$\odot$ and~$\oplus$ represent voxel-wise multiplication and addition, respectively. While~$\mathcal{L}_{sdt}$ and~$\mathcal{L}_{dice}$ both result in effective skull-stripping networks, we favor the distance loss~$\mathcal{L}_{sdt}$ due to its smoothing effect on the outline of the predicted brain mask, as demonstrated in Experiment~\ref{loss_comparison}.

\subsection{Implementation}

We implement~$g_\theta$ using a 3D U-Net convolutional architecture, with down-sampling (encoder) and up-sampling (decoder) components that facilitate the integration of features across large spatial regions. The U-Net comprises seven resolution levels, which each include two convolutional operations with leaky ReLU activations (parameter $\alpha=0.2$) and filter numbers defined in Figure~\ref{fig:training_schematic}. Down-sampling is achieved through max-pooling, and skip-connections are formed by concatenating the outputs of each encoder level with the inputs of the decoder level with corresponding resolution. In models using~$\mathcal{L} = \mathcal{L}_{sdt}$, one final, single-feature convolutional layer with linear activation outputs the predicted SDT~$d$. In models optimized with~$\mathcal{L} = \mathcal{L}_{dice}$, the final layer is a two-feature convolution, with softmax activation, that outputs a probabilistic segmentation representing non-brain and brain regions.

We train SynthStrip using the Adam optimizer~\citep{kingma2014adam} with a batch size of one and an initial learning rate of~$10^{-4}$. This rate is reduced by a factor of two after every 20,000 optimization steps without a decrease in validation loss. At test-time, all inputs to the model are internally conformed to~1-\textit{mm} isotropic voxel size using trilinear interpolation, and intensities are scaled between 0 and 1. The U-Net outputs are resampled such that the final brain mask is computed in the original input space. We implement SynthStrip in Python, using the open-source PyTorch~\citep{paszke2019} and Neurite~\citep{dalca2018anatomical} libraries, and make our tool and associated code available in the open-source FreeSurfer package (\url{https://w3id.org/synthstrip}). All experiments are conducted using Intel Xeon Silver 4214R CPUs and Nvidia RTX 8000 GPUs.

\subsection{Data}

In our experiments, we employ a small training dataset of adult and infant brain segmentations and a separate, larger dataset of acquired images for validation and testing that spans across age, health, resolution, and imaging modality. All data are 3D images, acquired either directly or as stacks of 2D MRI slices.

\subsubsection{Training Data}
\label{train_data}

\subpara{Datasets} We compose a set of 80 training subjects, each with whole-head tissue segmentations, from the following three cohorts: 40 adult subjects from the Buckner40 dataset~\citep{fischl2002}, 30 locally scanned adult subjects from the Human Connectome Aging Project (HCP-A)~\citep{bookheimer2019,harms2018}, and 10 infant subjects born full-term, scanned at Boston Children's Hospital at ages between 0 and 18 months~\citep{de2015}.

\subpara{Processing} To compute anatomical segmentations of individual cerebral regions, adult and infant T1w scans are processed with SAMSEG~\citep{puonti2016} and the Infant FreeSurfer reconstruction pipeline~\citep{zollei2020}, respectively. In order to build complete segmentation maps for robust whole-head image synthesis, we also generate six coarse labels of extra-cerebral tissue using a simple intensity-based labeling strategy with thresholds that mark label intensity boundaries. Considering only non-zero voxels without brain labels, we fit threshold values to each image by maximizing the similarity in number of voxels for each extra-cerebral label. These extra-cerebral labels do not necessarily represent or differentiate meaningful anatomical structures -- their purpose is to provide intensity and spatial variability to synthesized regions outside the brain.

In total, the training segmentations contain 46 individual anatomical labels, with 40 brain-specific labels (including CSF), that we merge into the target brain mask~$\hat{y}$. All training segmentations are fit to a~$256 ^ 3$ image shape with~1-\textit{mm} isotropic resolution. We emphasize that this geometric preprocessing is not required at test-time.

\subsubsection{Evaluation Data}
\label{evaluation_data}

\subpara{Datasets} Our evaluation data comprise 620 images, split into validation and test subsets of sizes 22 and 598, respectively. We gather these images across seven public datasets, with makeup, resolution, and validation splits outlined in Table~\ref{eval_datasets}. The IXI\footnote{Acquired from~\url{http://brain-development.org/ixi-dataset}.} dataset features a range of MRI contrasts and modalities, including T1w and T2w as well as PDw, MRA, and DWI. To simplify the DWI evaluation, a single diffusion direction is randomly extracted from each acquisition. The FSM subset~\citep{greve2021} is derived from in-house data using standard acquisitions as well as quantitative T1 maps (qT1). In-house, pseudo-continuous ASL~(PCASL) scans are acquired as stacks of 2D-EPI slices with low resolution and a small FOV that often crops the ventral brain region~\citep{dai2008continuous}. The QIN~\citep{clark2013cancer,qin,prah2015repeatability} dataset comprises pre-contrast, clinical stacks of thick image slices from patients with newly diagnosed glioblastoma. We also include a subset of the infant T1w image dataset, using subjects held-out from training. Lastly, to evaluate the ability of SynthStrip to adapt to imaging modalities beyond MR, we gather a test cohort of brain CT and FDG-PET scans from the CERMEP-IDB-MRXFDG (CIM) database~\citep{merida2021cermep}.

\begin{table}[t]
\centering
\small
\begin{tabular}{lllcc}
\toprule
Dataset & Modality  & Res.\ (\textit{mm}$^3$)       & Val.\ & Test \\
\hline \\ [-2.5ex]
IXI
& T1w MRI & $0.9{\times}0.9{\times}1.2$ & 0          & 48    \\
& T2w MRI & $0.9{\times}0.9{\times}1.2$ & 2          & 48    \\
& PDw MRI & $0.9{\times}0.9{\times}1.2$ & 2          & 48    \\
& MRA & $0.5{\times}0.5{\times}0.8$ & 2          & 48    \\
& DWI & $1.8{\times}1.8{\times}2.0$ & 0          & 32    \\
\hline \\ [-2.5ex]
FSM
& T1w MPRAGE  & $1.0{\times}1.0{\times}1.0$ & 0          & 38    \\
& T2w 3D-SPACE        & $1.0{\times}1.0{\times}1.0$ & 2          & 34    \\
& PDw 3D-FLASH         & $1.0{\times}1.0{\times}1.0$ & 2          & 30    \\
& qT1 MP2RAGE        & $1.0{\times}1.0{\times}1.0$ & 2          & 30    \\
\hline \\ [-2.5ex]
ASL
& T1w MPRAGE  & $1.0{\times}1.0{\times}1.0$ & 2          & 41    \\
& PCASL 2D-EPI & $3.4{\times}3.4{\times}5.0$ & 2          & 41    \\
\hline \\ [-2.5ex]
QIN
& T1w 2D-FLASH & $0.4{\times}0.4{\times}6.0$ & 2          & 52    \\
& T2-FLAIR 2D-FSE      & $0.4{\times}0.4{\times}6.0$ & 2          & 15    \\
& T2w 2D-FSE         & $1.0{\times}1.0{\times}5.0$ & 2          & 37    \\
\hline \\ [-2.5ex]
Infant
& T1w MPRAGE  & $1.0{\times}1.0{\times}1.0$ & 0          & 16    \\
\hline \\ [-2.5ex]
CIM
& FDG PET     & $2.0{\times}2.0{\times}2.0$ & 0          & 20    \\
& CT          & $0.6{\times}0.6{\times}1.5$ & 0          & 20    \\
\bottomrule
\end{tabular}
    \caption{We employ a diverse set of acquired evaluation data, spanning across imaging modalities, MRI contrasts, and resolution (res.), where 2D indicates stacks of slice-wise acquisitions. Each individual dataset is divided into a small validation (val.)\ and a larger test set. For further details see Section~\ref{evaluation_data}.}
\label{eval_datasets}
\end{table}

\subpara{Ground-truth masks} For each image in the evaluation dataset, we derive a reference brain mask using the following labelling strategy. Since every evaluation subject includes a corresponding T1w image, we generate brain masks for these scans using each \textit{classical} baseline method evaluated in our analysis. Then, an ``average'' brain mask is computed for each subject by extracting the majority label value at every voxel. We refine the average masks manually before propagating the masks by rigidly aligning each subject's T1w scan to the remaining image types with a robust registration approach~\citep{reuter2010}. Poor alignments are further refined by hand. We make the reference dataset available online to facilitate future development of skull-stripping techniques, including the original images if permitted by their respective licenses.

\begin{figure*}[t]
    \centering
    \includegraphics[width=\textwidth]{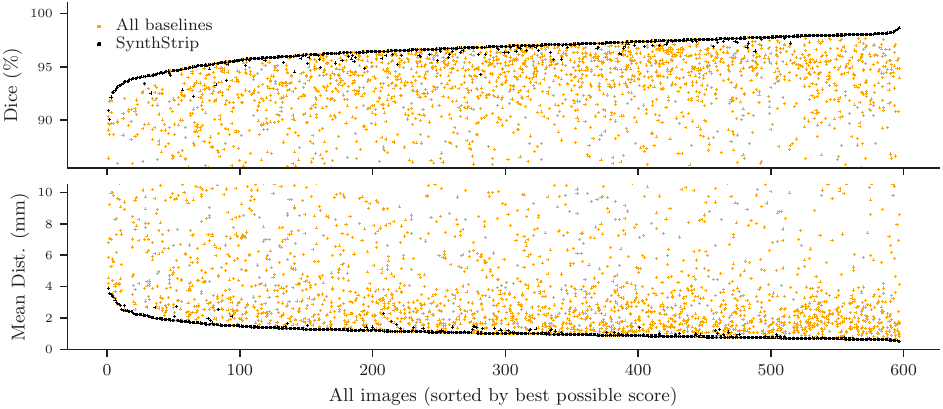}
    \caption{SynthStrip accuracy compared to baseline methods, across all images in the test set. Images are sorted by the score of the top performing skull-stripping method. Each dot represents a single brain mask derived with a particular tool, and each column of dots represents the scores obtained for a single image across tools.}
    \label{fig:all_subjects}
\end{figure*}

\section{Experiments}

We analyze the performance of SynthStrip on diverse whole-head images and compare its 3D skull-stripping accuracy to classical and deep-learning baseline tools.

\subpara{Baselines} We select a group of skull-stripping baselines based on their popularity, determined by citation count, and effectiveness, as shown in prior work~\citep{fatima2020state,iglesias2011robust}. As classical baselines, we choose ROBEX 1.1, BET from FSL 6.0.4, 3dSkullStrip (3DSS) from AFNI 21.0.21, BEaST 1.15, and the FreeSurfer 7.2 watershed algorithm (FSW). Unfortunately, many top-cited, learning-based approaches do not make their models available, even upon request to the authors. A notable exception is Deep MRI Brain Extraction (DMBE), which we therefore include. Default parameters are used for each method except BET, for which the~\texttt{-R} option is provided for more accurate brain center estimation. All inputs to FSW and DMBE are resampled to~1-\textit{mm} isotropic voxel sizes to accommodate the expected input resolution for these methods.

\subpara{Metrics} We evaluate the similarity between computed and ground-truth brain masks by measuring their Dice overlap, mean and maximum (Hausdorff) surface distances, and percent difference in total volume. Baseline scores are compared to SynthStrip with a paired sample t-test. Sensitivity and specificity, which measure the percent of true positive and true negative brain labels, respectively, provide further insight into the properties of the computed brain masks.

\begin{table*}[]
\centering
\begin{tabular}{lllllllll}

\multicolumn{8}{c}{\textbf{Mean Surface Distance (mm)}} \\
[1ex] \toprule \\ [-2ex]
                 &   SynthStrip  &  ROBEX  &  BET  &  3DSS  &  BEaST  &  FSW  &  DMBE  \\
[0.5ex] \toprule \\ [-2.5ex]
IXI T1w          &  $1.0 \pm 0.2$  &  $1.2 \pm 0.3$  &  $1.4 \pm 1.5$  &  $8.9 \pm 6.3$  &  $2.3 \pm 0.3$  &  $2.9 \pm 2.2$  &  $3.6 \pm 3.5$  \\
&    &  {\footnotesize $2.3 \times 10^{-6}$}  &  {\footnotesize {\color{orange} $1.1 \times 10^{-1}$}}  &  {\footnotesize $8.6 \times 10^{-12}$}  &  {\footnotesize $7.3 \times 10^{-33}$}  &  {\footnotesize $1.2 \times 10^{-7}$}  &  {\footnotesize $5.4 \times 10^{-6}$}  \\
\hline \\ [-2.3ex]
FSM T1w          &  $0.7 \pm 0.1$  &  $1.4 \pm 0.3$  &  $18.8 \pm 6.7$  &  $3.0 \pm 1.3$  &  $2.8 \pm 0.2$  &  $2.5 \pm 2.0$  &  $8.2 \pm 5.0$  \\
&    &  {\footnotesize $5.7 \times 10^{-17}$}  &  {\footnotesize $1.2 \times 10^{-18}$}  &  {\footnotesize $3.7 \times 10^{-12}$}  &  {\footnotesize $7.2 \times 10^{-36}$}  &  {\footnotesize $6.8 \times 10^{-6}$}  &  {\footnotesize $4.6 \times 10^{-11}$}  \\
\hline \\ [-2.3ex]
ASL T1w          &  $0.9 \pm 0.2$  &  $1.0 \pm 0.4$  &  $1.7 \pm 0.4$  &  $3.4 \pm 1.5$  &  $3.9 \pm 2.2$  &  $1.6 \pm 0.4$  &  $4.3 \pm 2.3$  \\
&    &  {\footnotesize $3.1 \times 10^{-2}$}  &  {\footnotesize $6.0 \times 10^{-16}$}  &  {\footnotesize $4.4 \times 10^{-14}$}  &  {\footnotesize $9.5 \times 10^{-12}$}  &  {\footnotesize $1.6 \times 10^{-11}$}  &  {\footnotesize $2.0 \times 10^{-12}$}  \\
\hline \\ [-2.3ex]
QIN T1w          &  $1.1 \pm 0.4$  &  $2.8 \pm 2.4$  &  $2.3 \pm 1.2$  &  $11.7 \pm 5.6$  &  $5.8 \pm 7.8$  &  $9.8 \pm 8.0$  &  $3.7 \pm 2.2$  \\
&    &  {\footnotesize $9.5 \times 10^{-7}$}  &  {\footnotesize $1.7 \times 10^{-11}$}  &  {\footnotesize $2.3 \times 10^{-19}$}  &  {\footnotesize $3.9 \times 10^{-5}$}  &  {\footnotesize $1.7 \times 10^{-10}$}  &  {\footnotesize $4.0 \times 10^{-12}$}  \\
\hline \\ [-2.3ex]
IXI T2w          &  $1.2 \pm 0.3$  &  $3.4 \pm 1.3$  &  $3.2 \pm 2.3$  &  $9.6 \pm 5.1$  &  $20.5 \pm 12.4$  &  -  &  $57.2 \pm 19.7$  \\
&    &  {\footnotesize $1.8 \times 10^{-17}$}  &  {\footnotesize $1.6 \times 10^{-7}$}  &  {\footnotesize $3.0 \times 10^{-15}$}  &  {\footnotesize $1.8 \times 10^{-14}$}  &  -  &  {\footnotesize $4.8 \times 10^{-25}$}  \\
\hline \\ [-2.3ex]
FSM T2w          &  $0.8 \pm 0.1$  &  $2.6 \pm 0.8$  &  $1.6 \pm 0.7$  &  $1.9 \pm 0.9$  &  $14.7 \pm 10.6$  &  -  &  $72.4 \pm 24.2$  \\
&    &  {\footnotesize $1.7 \times 10^{-15}$}  &  {\footnotesize $5.7 \times 10^{-8}$}  &  {\footnotesize $8.5 \times 10^{-9}$}  &  {\footnotesize $3.6 \times 10^{-9}$}  &  -  &  {\footnotesize $6.4 \times 10^{-19}$}  \\
\hline \\ [-2.3ex]
QIN T2w          &  $1.6 \pm 0.8$  &  $4.6 \pm 2.5$  &  $3.9 \pm 2.1$  &  $14.9 \pm 9.6$  &  $16.8 \pm 9.3$  &  -  &  $11.8 \pm 4.8$  \\
&    &  {\footnotesize $1.9 \times 10^{-9}$}  &  {\footnotesize $9.1 \times 10^{-8}$}  &  {\footnotesize $5.9 \times 10^{-9}$}  &  {\footnotesize $7.2 \times 10^{-12}$}  &  -  &  {\footnotesize $2.8 \times 10^{-16}$}  \\
\hline \\ [-2.3ex]
QIN FLAIR        &  $1.0 \pm 0.2$  &  $2.1 \pm 0.7$  &  $1.2 \pm 0.3$  &  $9.9 \pm 5.9$  &  $3.4 \pm 1.9$  &  $7.9 \pm 3.1$  &  $4.5 \pm 1.2$  \\
&    &  {\footnotesize $3.1 \times 10^{-5}$}  &  {\footnotesize $5.3 \times 10^{-3}$}  &  {\footnotesize $1.6 \times 10^{-5}$}  &  {\footnotesize $1.2 \times 10^{-4}$}  &  {\footnotesize $1.4 \times 10^{-7}$}  &  {\footnotesize $1.5 \times 10^{-9}$}  \\
\hline \\ [-2.3ex]
IXI PDw          &  $1.2 \pm 0.4$  &  $1.9 \pm 0.6$  &  $1.6 \pm 0.6$  &  $9.8 \pm 5.5$  &  $11.2 \pm 11.2$  &  $8.5 \pm 5.8$  &  $5.1 \pm 2.4$  \\
&    &  {\footnotesize $5.9 \times 10^{-15}$}  &  {\footnotesize $3.0 \times 10^{-7}$}  &  {\footnotesize $1.4 \times 10^{-14}$}  &  {\footnotesize $1.2 \times 10^{-7}$}  &  {\footnotesize $5.7 \times 10^{-11}$}  &  {\footnotesize $2.4 \times 10^{-15}$}  \\
\hline \\ [-2.3ex]
FSM PDw          &  $1.0 \pm 0.2$  &  $1.5 \pm 0.4$  &  $1.8 \pm 3.7$  &  $1.6 \pm 0.6$  &  $4.3 \pm 3.3$  &  $20.0 \pm 7.1$  &  $17.9 \pm 3.5$  \\
&    &  {\footnotesize $7.3 \times 10^{-6}$}  &  {\footnotesize {\color{orange} $2.3 \times 10^{-1}$}}  &  {\footnotesize $8.9 \times 10^{-7}$}  &  {\footnotesize $4.5 \times 10^{-6}$}  &  {\footnotesize $1.2 \times 10^{-15}$}  &  {\footnotesize $5.6 \times 10^{-23}$}  \\
\hline \\ [-2.3ex]
IXI MRA          &  $1.3 \pm 0.4$  &  $10.7 \pm 2.8$  &  $16.2 \pm 6.5$  &  $9.1 \pm 4.0$  &  $4.2 \pm 2.4$  &  -  &  $34.8 \pm 17.1$  \\
&    &  {\footnotesize $3.0 \times 10^{-28}$}  &  {\footnotesize $7.8 \times 10^{-21}$}  &  {\footnotesize $4.3 \times 10^{-18}$}  &  {\footnotesize $9.8 \times 10^{-11}$}  &  -  &  {\footnotesize $2.4 \times 10^{-18}$}  \\
\hline \\ [-2.3ex]
FSM qT1          &  $0.8 \pm 0.1$  &  $21.8 \pm 14.8$  &  $33.0 \pm 10.2$  &  $33.2 \pm 3.0$  &  $31.9 \pm 23.7$  &  $26.0 \pm 11.9$  &  $54.0 \pm 17.7$  \\
&    &  {\footnotesize $1.2 \times 10^{-8}$}  &  {\footnotesize $1.2 \times 10^{-17}$}  &  {\footnotesize $2.3 \times 10^{-32}$}  &  {\footnotesize $5.4 \times 10^{-8}$}  &  {\footnotesize $1.4 \times 10^{-12}$}  &  {\footnotesize $4.2 \times 10^{-17}$}  \\
\hline \\ [-2.3ex]
ASL EPI          &  $1.6 \pm 0.6$  &  $4.8 \pm 2.3$  &  $2.0 \pm 0.7$  &  $9.6 \pm 4.0$  &  $14.8 \pm 11.8$  &  $16.1 \pm 5.4$  &  $2.6 \pm 0.6$  \\
&    &  {\footnotesize $1.6 \times 10^{-10}$}  &  {\footnotesize $9.9 \times 10^{-5}$}  &  {\footnotesize $9.8 \times 10^{-16}$}  &  {\footnotesize $9.8 \times 10^{-9}$}  &  {\footnotesize $5.0 \times 10^{-20}$}  &  {\footnotesize $2.5 \times 10^{-15}$}  \\
\hline \\ [-2.3ex]
Infant T1w       &  $1.0 \pm 0.3$  &  $4.1 \pm 6.1$  &  $14.3 \pm 10.7$  &  $22.2 \pm 12.1$  &  $19.0 \pm 21.7$  &  $17.6 \pm 15.5$  &  $6.6 \pm 3.4$  \\
&    &  {\footnotesize {\color{orange} $5.9 \times 10^{-2}$}}  &  {\footnotesize $2.1 \times 10^{-4}$}  &  {\footnotesize $5.3 \times 10^{-6}$}  &  {\footnotesize $5.5 \times 10^{-3}$}  &  {\footnotesize $1.3 \times 10^{-3}$}  &  {\footnotesize $1.6 \times 10^{-5}$}  \\
\hline \\ [-2.3ex]
IXI DWI          &  $1.6 \pm 0.6$  &  $6.2 \pm 3.0$  &  $2.4 \pm 0.9$  &  $8.5 \pm 2.7$  &  $11.1 \pm 9.3$  &  $11.2 \pm 4.7$  &  $6.8 \pm 1.2$  \\
&    &  {\footnotesize $2.5 \times 10^{-9}$}  &  {\footnotesize $2.7 \times 10^{-11}$}  &  {\footnotesize $9.8 \times 10^{-15}$}  &  {\footnotesize $2.5 \times 10^{-6}$}  &  {\footnotesize $2.3 \times 10^{-12}$}  &  {\footnotesize $5.1 \times 10^{-27}$}  \\
\hline \\ [-2.3ex]
CIM PET          &  $1.5 \pm 0.4$  &  $3.9 \pm 2.2$  &  $9.3 \pm 4.0$  &  $2.2 \pm 2.9$  &  $69.0 \pm 21.2$  &  $16.2 \pm 3.5$  &  $17.6 \pm 6.4$  \\
&    &  {\footnotesize $7.7 \times 10^{-5}$}  &  {\footnotesize $8.0 \times 10^{-8}$}  &  {\footnotesize {\color{orange} $2.4 \times 10^{-1}$}}  &  {\footnotesize $2.6 \times 10^{-10}$}  &  {\footnotesize $5.8 \times 10^{-13}$}  &  {\footnotesize $1.0 \times 10^{-9}$}  \\
\hline \\ [-2.3ex]
CIM CT           &  $2.0 \pm 0.4$  &  $11.4 \pm 1.2$  &  $34.1 \pm 3.7$  &  $20.6 \pm 2.0$  &  $74.8 \pm 18.4$  &  $29.1 \pm 5.9$  &  $34.7 \pm 8.2$  \\
&    &  {\footnotesize $4.1 \times 10^{-20}$}  &  {\footnotesize $3.6 \times 10^{-19}$}  &  {\footnotesize $4.1 \times 10^{-20}$}  &  {\footnotesize $7.7 \times 10^{-12}$}  &  {\footnotesize $2.9 \times 10^{-12}$}  &  {\footnotesize $5.5 \times 10^{-13}$}  \\
\bottomrule \\ [0ex]

\end{tabular}
\caption{SynthStrip and baseline method accuracy across datasets, as measured by the mean surface distance ($\pm$~SD) between computed and ground-truth binary brain masks.~\textit{p}-values comparing baseline with SythStrip results are presented below each score. SynthStrip stands out as a dominant skull-stripping technique, significantly outperforming baselines for nearly every dataset with the exception of those with~\textit{p}-values in orange, for which~\textit{p}~$>$ 0.05. FSW fails entirely for multiple subsets of non-T1w images (metrics not shown).}
\label{table_dist_mean}
\end{table*}

\begin{table*}[]
\centering
\begin{tabular}{lllllllll}

\multicolumn{8}{c}{\textbf{Dice (\%)}} \\
[1ex] \toprule \\ [-2ex]
                 &   SynthStrip  &  ROBEX  &  BET  &  3DSS  &  BEaST  &  FSW  &  DMBE  \\
[0.5ex] \toprule \\ [-2.5ex]
IXI T1w          &  $97.0 \pm 0.5$  &  $96.2 \pm 0.8$  &  $96.1 \pm 3.1$  &  $95.4 \pm 1.4$  &  $93.4 \pm 0.8$  &  $92.6 \pm 4.4$  &  $93.7 \pm 3.0$  \\
&    &  {\footnotesize $4.5 \times 10^{-8}$}  &  {\footnotesize {\color{orange} $5.1 \times 10^{-2}$}}  &  {\footnotesize $1.3 \times 10^{-9}$}  &  {\footnotesize $1.1 \times 10^{-32}$}  &  {\footnotesize $5.8 \times 10^{-9}$}  &  {\footnotesize $8.0 \times 10^{-10}$}  \\
\hline \\ [-2.3ex]
FSM T1w          &  $97.8 \pm 0.3$  &  $95.9 \pm 0.7$  &  $65.8 \pm 11.2$  &  $92.0 \pm 3.0$  &  $92.1 \pm 0.7$  &  $93.8 \pm 3.7$  &  $89.5 \pm 3.3$  \\
&    &  {\footnotesize $3.3 \times 10^{-18}$}  &  {\footnotesize $1.7 \times 10^{-19}$}  &  {\footnotesize $1.4 \times 10^{-13}$}  &  {\footnotesize $1.9 \times 10^{-35}$}  &  {\footnotesize $2.2 \times 10^{-7}$}  &  {\footnotesize $5.6 \times 10^{-18}$}  \\
\hline \\ [-2.3ex]
ASL T1w          &  $97.3 \pm 0.5$  &  $96.8 \pm 1.1$  &  $94.9 \pm 1.1$  &  $90.8 \pm 3.5$  &  $89.0 \pm 5.5$  &  $95.5 \pm 0.9$  &  $92.7 \pm 1.7$  \\
&    &  {\footnotesize $3.0 \times 10^{-3}$}  &  {\footnotesize $5.7 \times 10^{-17}$}  &  {\footnotesize $1.5 \times 10^{-15}$}  &  {\footnotesize $7.7 \times 10^{-13}$}  &  {\footnotesize $1.1 \times 10^{-11}$}  &  {\footnotesize $1.1 \times 10^{-19}$}  \\
\hline \\ [-2.3ex]
QIN T1w          &  $96.3 \pm 0.8$  &  $92.8 \pm 4.6$  &  $93.8 \pm 2.7$  &  $92.4 \pm 3.1$  &  $85.2 \pm 17.1$  &  $79.9 \pm 13.3$  &  $89.0 \pm 7.3$  \\
&    &  {\footnotesize $3.7 \times 10^{-7}$}  &  {\footnotesize $5.2 \times 10^{-11}$}  &  {\footnotesize $2.3 \times 10^{-15}$}  &  {\footnotesize $1.5 \times 10^{-5}$}  &  {\footnotesize $5.2 \times 10^{-12}$}  &  {\footnotesize $7.4 \times 10^{-10}$}  \\
\hline \\ [-2.3ex]
IXI T2w          &  $96.4 \pm 0.7$  &  $91.3 \pm 2.7$  &  $91.0 \pm 6.2$  &  $94.9 \pm 1.7$  &  $63.0 \pm 12.4$  &  -  &  $6.7 \pm 8.0$  \\
&    &  {\footnotesize $2.9 \times 10^{-19}$}  &  {\footnotesize $2.0 \times 10^{-7}$}  &  {\footnotesize $1.4 \times 10^{-9}$}  &  {\footnotesize $7.8 \times 10^{-24}$}  &  -  &  {\footnotesize $1.1 \times 10^{-52}$}  \\
\hline \\ [-2.3ex]
FSM T2w          &  $97.7 \pm 0.3$  &  $93.2 \pm 1.7$  &  $95.7 \pm 1.6$  &  $94.9 \pm 2.0$  &  $69.8 \pm 15.0$  &  -  &  $3.4 \pm 7.2$  \\
&    &  {\footnotesize $4.9 \times 10^{-18}$}  &  {\footnotesize $9.8 \times 10^{-9}$}  &  {\footnotesize $4.1 \times 10^{-10}$}  &  {\footnotesize $7.3 \times 10^{-13}$}  &  -  &  {\footnotesize $1.1 \times 10^{-40}$}  \\
\hline \\ [-2.3ex]
QIN T2w          &  $95.2 \pm 1.1$  &  $87.3 \pm 5.3$  &  $89.6 \pm 4.0$  &  $71.4 \pm 21.2$  &  $57.7 \pm 19.1$  &  -  &  $61.5 \pm 16.5$  \\
&    &  {\footnotesize $1.1 \times 10^{-10}$}  &  {\footnotesize $1.7 \times 10^{-9}$}  &  {\footnotesize $6.3 \times 10^{-7}$}  &  {\footnotesize $9.1 \times 10^{-14}$}  &  -  &  {\footnotesize $2.9 \times 10^{-15}$}  \\
\hline \\ [-2.3ex]
QIN FLAIR        &  $96.4 \pm 0.5$  &  $93.7 \pm 1.3$  &  $95.9 \pm 0.9$  &  $93.8 \pm 1.1$  &  $90.3 \pm 4.7$  &  $83.4 \pm 6.2$  &  $87.9 \pm 3.0$  \\
&    &  {\footnotesize $5.7 \times 10^{-7}$}  &  {\footnotesize $1.3 \times 10^{-2}$}  &  {\footnotesize $4.2 \times 10^{-9}$}  &  {\footnotesize $8.7 \times 10^{-5}$}  &  {\footnotesize $2.6 \times 10^{-7}$}  &  {\footnotesize $1.5 \times 10^{-9}$}  \\
\hline \\ [-2.3ex]
IXI PDw          &  $96.4 \pm 1.0$  &  $94.6 \pm 1.3$  &  $95.5 \pm 1.4$  &  $95.1 \pm 1.6$  &  $78.3 \pm 15.8$  &  $81.3 \pm 10.9$  &  $90.0 \pm 3.8$  \\
&    &  {\footnotesize $5.3 \times 10^{-17}$}  &  {\footnotesize $1.9 \times 10^{-7}$}  &  {\footnotesize $1.4 \times 10^{-8}$}  &  {\footnotesize $1.8 \times 10^{-10}$}  &  {\footnotesize $2.5 \times 10^{-12}$}  &  {\footnotesize $3.8 \times 10^{-16}$}  \\
\hline \\ [-2.3ex]
FSM PDw          &  $97.2 \pm 0.5$  &  $95.8 \pm 1.0$  &  $95.7 \pm 6.5$  &  $95.5 \pm 1.4$  &  $88.8 \pm 8.4$  &  $67.4 \pm 9.8$  &  $79.1 \pm 4.4$  \\
&    &  {\footnotesize $1.1 \times 10^{-6}$}  &  {\footnotesize {\color{orange} $2.3 \times 10^{-1}$}}  &  {\footnotesize $4.7 \times 10^{-8}$}  &  {\footnotesize $5.1 \times 10^{-6}$}  &  {\footnotesize $4.9 \times 10^{-17}$}  &  {\footnotesize $3.4 \times 10^{-21}$}  \\
\hline \\ [-2.3ex]
IXI MRA          &  $97.7 \pm 0.5$  &  $82.0 \pm 4.3$  &  $62.3 \pm 14.9$  &  $95.1 \pm 1.0$  &  $91.5 \pm 5.3$  &  -  &  $18.1 \pm 19.5$  \\
&    &  {\footnotesize $4.5 \times 10^{-30}$}  &  {\footnotesize $1.3 \times 10^{-21}$}  &  {\footnotesize $3.0 \times 10^{-24}$}  &  {\footnotesize $6.7 \times 10^{-10}$}  &  -  &  {\footnotesize $4.7 \times 10^{-32}$}  \\
\hline \\ [-2.3ex]
FSM qT1         &  $97.7 \pm 0.3$  &  $63.1 \pm 19.6$  &  $47.2 \pm 13.5$  &  $44.6 \pm 3.8$  &  $36.3 \pm 15.6$  &  $49.5 \pm 10.7$  &  $3.1 \pm 7.1$  \\
&    &  {\footnotesize $1.1 \times 10^{-10}$}  &  {\footnotesize $8.8 \times 10^{-20}$}  &  {\footnotesize $4.9 \times 10^{-36}$}  &  {\footnotesize $9.4 \times 10^{-20}$}  &  {\footnotesize $1.7 \times 10^{-21}$}  &  {\footnotesize $1.6 \times 10^{-36}$}  \\
\hline \\ [-2.3ex]
ASL EPI          &  $95.2 \pm 1.4$  &  $88.3 \pm 4.6$  &  $94.2 \pm 1.8$  &  $95.2 \pm 1.4$  &  $67.4 \pm 19.6$  &  $69.4 \pm 11.9$  &  $92.8 \pm 1.5$  \\
&    &  {\footnotesize $3.6 \times 10^{-11}$}  &  {\footnotesize $5.2 \times 10^{-4}$}  &  {\footnotesize {\color{orange} $5.8 \times 10^{-1}$}}  &  {\footnotesize $2.0 \times 10^{-11}$}  &  {\footnotesize $8.9 \times 10^{-17}$}  &  {\footnotesize $2.1 \times 10^{-15}$}  \\
\hline \\ [-2.3ex]
Infant T1w       &  $96.1 \pm 1.4$  &  $87.4 \pm 15.4$  &  $66.8 \pm 20.3$  &  $71.6 \pm 17.7$  &  $63.2 \pm 38.4$  &  $61.3 \pm 31.3$  &  $84.4 \pm 7.3$  \\
&    &  {\footnotesize $3.5 \times 10^{-2}$}  &  {\footnotesize $3.0 \times 10^{-5}$}  &  {\footnotesize $4.4 \times 10^{-5}$}  &  {\footnotesize $3.9 \times 10^{-3}$}  &  {\footnotesize $7.9 \times 10^{-4}$}  &  {\footnotesize $4.9 \times 10^{-6}$}  \\
\hline \\ [-2.3ex]
IXI DWI          &  $95.5 \pm 1.3$  &  $85.3 \pm 5.7$  &  $93.4 \pm 2.1$  &  $90.4 \pm 2.3$  &  $79.5 \pm 11.4$  &  $75.1 \pm 8.8$  &  $80.5 \pm 3.1$  \\
&    &  {\footnotesize $2.4 \times 10^{-10}$}  &  {\footnotesize $1.2 \times 10^{-11}$}  &  {\footnotesize $1.6 \times 10^{-17}$}  &  {\footnotesize $6.8 \times 10^{-9}$}  &  {\footnotesize $1.5 \times 10^{-13}$}  &  {\footnotesize $1.4 \times 10^{-27}$}  \\
\hline \\ [-2.3ex]
CIM PET          &  $95.4 \pm 1.1$  &  $89.5 \pm 4.8$  &  $78.3 \pm 8.3$  &  $93.7 \pm 6.1$  &  $4.6 \pm 8.8$  &  $51.6 \pm 10.1$  &  $42.3 \pm 14.3$  \\
&    &  {\footnotesize $1.8 \times 10^{-5}$}  &  {\footnotesize $2.9 \times 10^{-8}$}  &  {\footnotesize {\color{orange} $2.1 \times 10^{-1}$}}  &  {\footnotesize $1.4 \times 10^{-18}$}  &  {\footnotesize $4.1 \times 10^{-13}$}  &  {\footnotesize $1.6 \times 10^{-12}$}  \\
\hline \\ [-2.3ex]
CIM CT           &  $94.3 \pm 0.9$  &  $73.7 \pm 2.0$  &  $45.4 \pm 4.0$  &  $60.4 \pm 2.9$  &  $2.3 \pm 3.3$  &  $48.8 \pm 5.1$  &  $58.2 \pm 6.5$  \\
&    &  {\footnotesize $3.8 \times 10^{-23}$}  &  {\footnotesize $4.8 \times 10^{-22}$}  &  {\footnotesize $4.0 \times 10^{-22}$}  &  {\footnotesize $3.5 \times 10^{-26}$}  &  {\footnotesize $1.4 \times 10^{-16}$}  &  {\footnotesize $2.1 \times 10^{-15}$}  \\
\bottomrule \\ [0ex]

\end{tabular}
\caption{Skull-stripping accuracy across datasets, as measured by the mean Dice overlap ($\pm$~SD) between computed and ground-truth binary brain masks.~\textit{p}-values comparing baseline with SythStrip results are presented below each score. Across each dataset, SynthStrip significantly outperforms most baselines except those with~\textit{p}-values in orange, for which~\textit{p}~$>$ 0.05.}
\label{table_dice}
\end{table*}

\begin{figure*}[t]
    \centering
    \textbf{T1w Near-Isotropic Evaluation Images}\par\medskip
    \includegraphics[width=\textwidth]{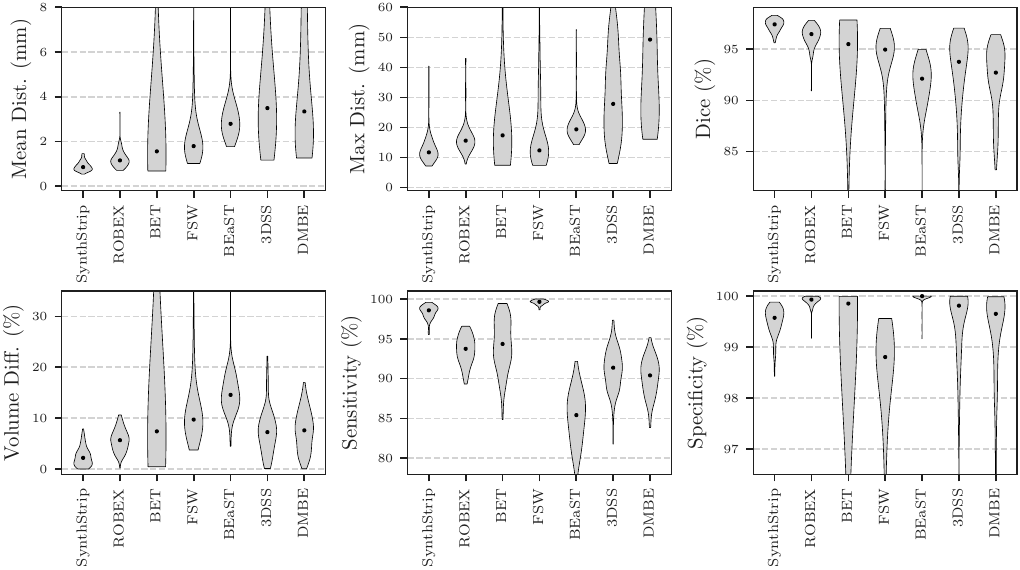}
    \caption{SynthStrip and baseline skull-stripping performance for near-isotropic, T1w adult MR brain images. Median scores are represented by black dots. For all metrics except sensitivity and specificity, SynthStrip yields optimal brain masks. The high specificity achieved by ROBEX and BEaST comes at the cost of substantial under-segmentation of the brain mask, as indicated by their low sensitivity scores. The inverse is true for FSW, which tends to substantially over-segment the brain. Black dots indicate median scores.}
    \label{fig:t1w_stats}
\end{figure*}

\begin{figure*}[t]
    \centering
    \textbf{Non-T1w, Thick-Slice, and Infant Evaluation Images}\par\medskip
    \includegraphics[width=\textwidth]{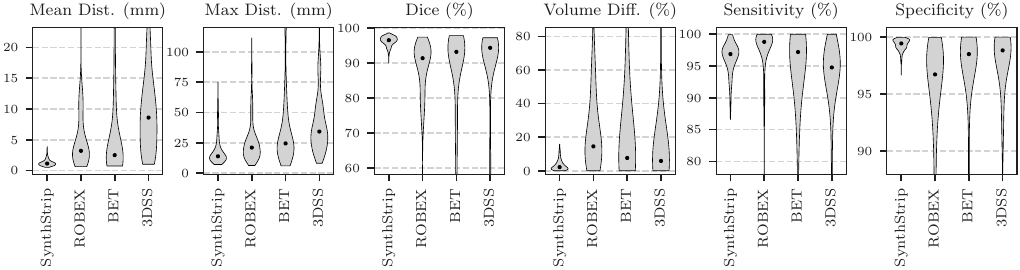}
    \caption{Considering all non-T1w, thick-slice, and infant images in the evaluation set, SynthStrip surpasses baseline accuracy by a wide margin. In this figure, we include only baselines that generalize to acquisition protocols and modalities beyond the common structural T1w MRI scans. Black dots indicate median scores.}
    \label{fig:non_t1_stats}
\end{figure*}

\subsection{Skull-Stripping Accuracy}

We assess the broad skull-stripping capability of a SynthStrip model trained using images synthesized from the label maps outlined in Section~\ref{train_data}. We compare the accuracy of our method to each of the baselines across the test set of real brain images defined in Section~\ref{evaluation_data}. Method runtime is compared for the FSM dataset.

The comparison demonstrates SynthStrip's accurate and robust brain extraction, which substantially outperforms baseline methods (Tables~\ref{table_dist_mean},~\ref{table_dice} and Supplementary Tables~\ref{table_dist_max},~\ref{table_voldiff}). For every evaluation metric, brain masks predicted by SynthStrip yield significantly better scores than baseline masks (\textit{p}~$<$ 0.05) for the~\textit{vast} majority of datasets. Importantly, no baseline method significantly outperforms SynthStrip on any dataset. As shown in Figure~\ref{fig:all_subjects}, SynthStrip achieves the highest Dice score \textit{and} lowest mean surface distance for more than 80\% of all test images, in stark contrast to the next best performing method, BET, which yields the top result for less than 10\% of images. The superior performance of SynthStrip persists even when considering only T1w, near-isotropic, adult-brain images, which all of the baselines are tuned for. Across this particular subset of 127 T1w images from the IXI, FSM, and ASL datasets, SynthStrip achieves the best mean Dice, surface distance, Hausdorff distance, and volume difference (Figure~\ref{fig:t1w_stats}), and it consistently extracts the brain with high specificity and sensitivity, while other methods tend to under-perform in either of those metrics due to tendencies to substantially over- or under-label the brain. When considering the remaining non-T1w, thick-slice, and infant image types, SynthStrip's predominance is similarly substantial (Figure~\ref{fig:non_t1_stats}). For FSM T1w data, our method runs on the CPU in less than one minute (Table~\ref{table_runtime}), trailing the fastest two baselines, BET and FSW, by approximately 17 seconds on average. On the GPU, SynthStrip runs substantially faster, requiring only~$1.8 \pm 0.2$ seconds.

\begin{table*}[b]
\centering
\begin{tabular}{llllllll}
\multicolumn{7}{c}{\textbf{CPU Runtime (minutes)}} \\
[1ex] \hline \\ [-2ex]
SynthStrip  &  ROBEX  &  BET  &  3DSS  &  BEaST  &  FSW  &  DMBE \\
\hline \\ [-2ex]
$0.48 \pm 0.01$ & $2.45 \pm 0.11$ & $0.20 \pm 0.14$ & $2.27 \pm 1.12$ & $4.14 \pm 0.24$ & $0.19 \pm 0.01$ & $48.89 \pm 4.72$ \\
\bottomrule \\ [0ex]
\end{tabular}
\caption{Average single-threaded CPU runtime ($\pm$SD) for T1w images from the FSM dataset. In addition to BET and FSW, SynthStrip is one of only three skull-stripping methods that consistently runs in under one minute.}
\label{table_runtime}
\end{table*}

\subsection{Qualitative Brain-Mask Analysis}

Across the evaluation set, skull-stripping errors in SynthStrip predictions are uncommon and typically involve minimally over-segmenting the brain mask by including thin regions of extra-cerebral matter near the dorsal cortex or pockets of tissue around the eye sockets, as shown in Figure 8. Considering only the $N$ images for which SynthStrip does not achieve the best score in Figure~\ref{fig:all_subjects}, on average, SynthStrip lags behind the best-performing baseline by  only~$-0.53 \pm 0.54$ Dice percentage points ($N = 111$) and~$(0.20 \pm 0.18)~mm$ mean surface distance ($N = 94$).

The top performing baseline method is ROBEX, which yields high-quality brain extraction across many of the test datasets, with the notable exception of the qT1 cohort. ROBEX produces spatially plausible brain masks and evades drastic failure modes that exist in other baselines, similarly to SynthStrip. However, despite its generally good performance, ROBEX has a tendency to include pockets of tissue surrounding the eyes and remove regions of cortical gray matter near the superior surface (Figures~\ref{fig:failures} and~\ref{fig:failure_grid}).

BET and 3DSS also perform effective brain extraction across image types, but tend to fail dramatically for outlier cases. For example, BET locates the brain boundary with considerable precision when successful. However, for some image subsets, especially those with abundant non-brain matter, such as FSM, BET often includes large regions of inferior skull as well as facial and neck tissue in the brain mask. While 3DSS largely avoids such gross mislabeling, it tends to produce skull-strips that leak into neck tissue or, conversely, remove small regions of the cortical surface.

BEaST and FSW perform well for near-isotropic T1w images, such as those in the IXI, FSM, and ASL datasets. But since they are heavily optimized for the assumed spatial and intensity features of this acquisition type, they generally perform poorly or even fail completely for other contrasts. Common error modes of FSW involve the failure to remove bits of skull or inferior non-brain matter, in contrast to BEaST, which is susceptible to removing critical regions of the cortex.

The learning-based method DMBE yields suitable brain masks for near-isotropic image types with T1w contrast but frequently leaves substantial, unconnected components of non-brain matter. While DMBE extracts the brain tissue border as opposed to CSF, our analysis shows that the predominant contributor to the discrepancy between DMBE and ground-truth brain masks is the inclusion of neck and facial tissue (Figures~\ref{fig:failures} and~\ref{fig:failure_grid}). DMBE model inference is slow, consuming more than a half hour to skull-strip a standard image.

\subsection{Variability Across Time-Series Data}
\label{intra_subject}

\begin{figure}[t]
    \centering
    \includegraphics[width=\columnwidth]{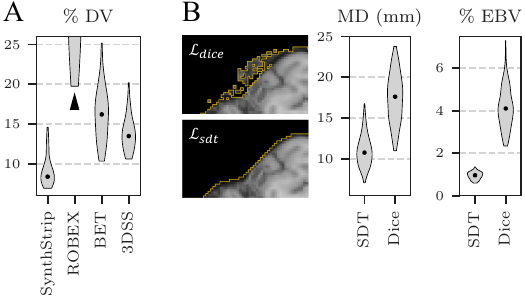}
    \caption{\textbf{A:} SynthStrip variability across time-series data, measured by percent of discordant voxel locations (DV) across diffusion-encoded directions, relative to the brain mask volume. The ROBEX median \% DV extends beyond the chart axis, as indicated by the black arrow. \textbf{B:} Effect of SDT- and Dice-based loss functions during training. A SynthStrip model trained using~$\mathcal{L}_{sdt}$ predicts substantially smoother brain masks (boundaries indicated in orange) than a model trained with~$\mathcal{L}_{dice}$, resulting in considerably lower maximum surface distance (MD) to ground truth masks and percent of exposed boundary voxels (EBV).
    }
    \label{fig:sd_and_loss}
\end{figure}

We analyze the consistency of SynthStrip brain masks across time-domain data by assessing the differences between diffusion-encoded directions acquired in the same session. For each subject in the DWI dataset, we affinely align and skull-strip all of the 16 diffusion-encoded frames in a common, average space~\citep{reuter2010}. We compute the number of discordant voxels across brain masks for a given method, defining discordant voxels (DV) as voxel locations with labels that differ in the time domain. We report the percent of DV relative to the brain mask volume, determined by the number of voxels labeled as brain in any frame. In this particular analysis, we only consider ROBEX, BET, and 3DSS as baselines since they generalize to DWI acquisitions. As shown in Figure~\ref{fig:sd_and_loss}, SynthStrip demonstrates a high level of intra-subject consistency, as it predicts brain masks with substantially lower \% DV across DWI directions than the baselines ($p < 10^{-12}$). Since the \% DV metric considers voxels labeled as brain for any direction, a single mask with gross mislabeling will substantially increase the metric value, as is the case with ROBEX, which over-segments the brain for only a few directions per subject.

\subsection{Loss Comparison}
\label{loss_comparison}

During our experimentation, we find that training SynthStrip models using a traditional soft Dice loss yields comparable results to those trained with an SDT-based loss for~\textit{nearly} every metric. However, despite similar global accuracy, we observe that models trained with~$\mathcal{L}_{dice}$ predict brain masks characterized by relatively noisy and rough boundaries, as illustrated in Figure~\ref{fig:sd_and_loss}. The high variability at the edge of the brain mask is emphasized by a~$6.4 \pm 3.2$~\textit{mm} increase in maximum surface distance when using~$\mathcal{L}_{dice}$ compared to~$\mathcal{L}_{sdt}$. We further quantify this discrepancy in brain-mask smoothness by computing the percent of exposed boundary voxels (EBV) that neighbor more non-brain labels than brain labels. Brain masks with noisier boundaries will exhibit larger EBV due to an increased mask surface area and number of sporadic border voxels. We perform this evaluation using the FSM data subset of 132 images with isotropic voxel size. Models trained with~$\mathcal{L}_{dice}$ predict masks with~$4.5\times$ higher EBV than models trained with~$\mathcal{L}_{sdt}$. We hypothesize that as the network learns to estimate an SDT, it is encouraged to focus more on the boundary of mask, rather than the label as a whole, resulting in a smoother prediction of the brain border.

\begin{figure}[!t]
    \centering
    \includegraphics[width=\columnwidth]{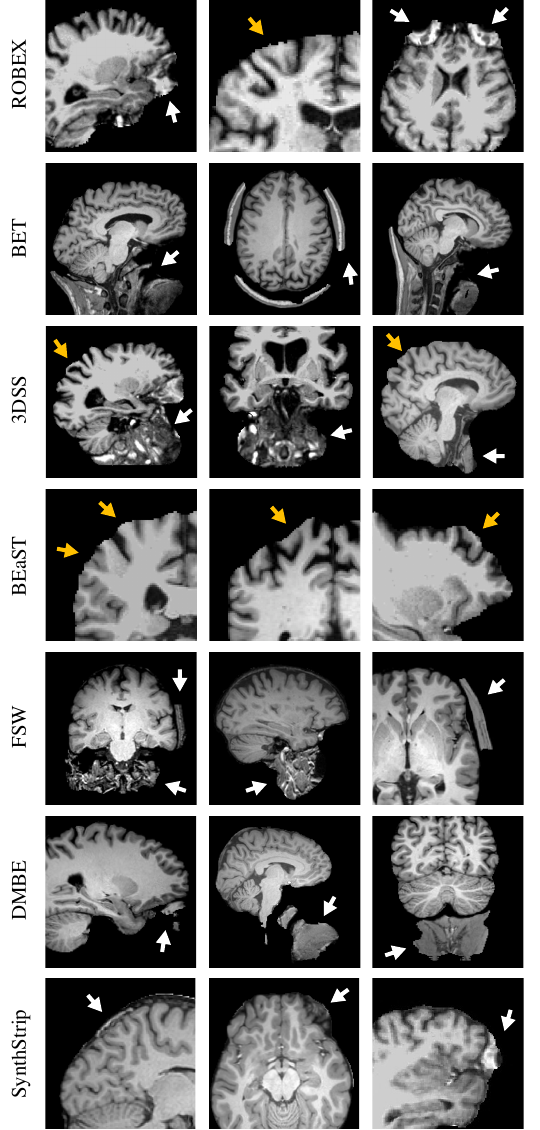}
    \caption{Representative skull-stripping errors for SynthStrip and baseline methods. White arrows indicate over-labeling of the brain mask, while orange arrows indicate removal of brain matter. SynthStrip errors are uncommon and typically involve including small regions of dura or other extracerebral tissue in the brain mask, if they occur.}
    \label{fig:failures}
\end{figure}

\section{Discussion}

We present SynthStrip, a learning-based, universal brain-extraction tool trained on diverse synthetic images. Subjected to training data that far exceeds the realistic range of medical images, the model learns to generalize across imaging modalities, anatomical variability, and acquisition schemes.

\subsection{Baseline Comparison}

SynthStrip significantly improves upon baseline skull-stripping accuracy for nearly every image cohort tested, and the few exceptions to this improvement involve data subsets for which SynthStrip matches baseline performance. This predominance is in part due to the ability of SynthStrip to generalize across a wide variety of image types as well as its proclivity to avoid substantial mislabeling. In particular, varying specific acquisition characteristics during synthesis promotes network robustness to such characteristics across a range of protocols. For example, simulating partial-volume effects with blurring and randomizing the resolution enable SynthStrip to accurately generalize to clinical thick-slice acquisitions and those with large voxel sizes. By learning robust, large-scale spatial features of representative brain masks, the model consistently predicts masks of realistic and expected shape. Baseline techniques, on the other hand, often rely on weak spatial priors and are therefore prone to over- or under-segment brain tissue when confronted with image features that are unexpected or unaccounted for (Figures~\ref{fig:failures} and~\ref{fig:failure_grid}).

ROBEX's consistent performance across contrasts and modality is somewhat unexpected since the discriminative edge detector is trained only for T1w scans. We hypothesize that the coupled shape model is able to compensate for any intensity bias encoded in the discriminative detector. The T1w-specific approaches BEAST and FSW could be effective for other MRI contrasts if provided known intensity priors of the brain matter. However, this work would require substantial human effort as it needs to be repeated for every new image type. The substantial, unconnected components of non-brain matter frequently left by DMBE are likely a byproduct of its convolutional architecture, which does not leverage multiple resolution levels to gather spatial features across large distances.

\subsection{Use for Brain-Specific Registration}

Consistent brain extraction across different images from the same subject is critical for accurate analysis of time-series acquisitions. For example, diffusion~\citep{holdsworth2012diffusion,jones2011diffusion} and functional MRI analyses~\citep{ashburner2009preparing,jenkinson2002improved} depend on within-subject registration of individual frames acquired across time to undo the effect of any head motion during the scan. Unfortunately, anatomical structures that deform non-rigidly between frames, such as the neck or tongue, can hamper brain-registration accuracy and thus impinge on downstream results. While this effect can be accounted for by first removing non-brain tissue from each frame to achieve brain-specific registration~\citep{andrade2018practical,fischmeister2013benefits}, it requires consistent brain extraction across frames~\citep{andrade2018practical,fein2006statistical,fischmeister2013benefits,hoffmann2020real}. SynthStrip's high within-subject consistency despite substantial contrast differences across the diffusion encoding demonstrates its potential for regularizing retrospective motion correction of time-series data.

\subsection{Model and Data Availability}

Even as learning-based methods in neuroimaging analysis continue to grow in popularity, developers of deep-learning skull-stripping tools are sometimes disinclined to provide easy-to-use distributions of their work. Out of the three promising methods discussed in this work, only DMBE makes its models and code publicly available for use. In contrast, we make SynthStrip available as a universal, cross-platform command-line utility, distributed both as a standalone and as a built-in FreeSurfer tool. To facilitate further development and testing of robust skull-stripping tools, we also make our evaluation data and ground-truth labels available at~\url{https://w3id.org/synthstrip}.

\subsection{Future Work}

While SynthStrip facilitates state-of-the-art brain extraction, we aim to extend the tissue-extraction strategy to other applications both within and beyond neuroimaging. One such application is fetal head extraction from in-utero fetal MRI scans. Due to excessive motion, fetal MRI is limited to the acquisition of sub-second 2D slices. However, stacks of several slices are needed to cover the anatomy of interest, and while their in-plane resolution is typically of the order of~1~\textit{mm} $\times$ 1~\textit{mm}, views across slices are hampered by slice thicknesses of 4-6~\textit{mm} and between-slice motion~\citep{hoffmann2021rapid}. To enable full 3D views of the fetal brain, post-processing tools for super-resolution reconstruction have emerged, that aim to reconstruct a high-quality volume of isotropic resolution from a number of slice stacks acquired at different angles~\citep{rousseau2006registration,kainz2015fast,ebner2020automated,iglesias2021joint}. Yet, these methods hinge on successful brain extraction which is challenging due to frequent artifacts and because the relatively small brain first needs to be localized within a wide FOV encompassing the maternal anatomy~\citep{gaudfernau2021analysis}. In addition, substantially fewer public fetal datasets are available for training in comparison to vast public adult brain datasets. This presents an ideal problem to be addressed with SynthStrip, as our approach synthesizes an endless stream of training data from only a handful of label maps.

\section{Conclusion}

The removal of non-brain signal from neuroimaging data is a fundamental first step for many quantitative analyses and its accuracy has a direct impact on downstream results. However, popular skull-stripping utilities are typically tailored to isotropic T1w scans and tend to fail, sometimes catastrophically, on images with other MRI contrasts or stack-of-slices acquisitions that are common in the clinic. We propose SynthStrip, a flexible tool that produces highly accurate brain masks across a landscape of imaging paradigms with widely varying contrast and resolution. We implement our method by leveraging anatomical label maps to synthesize a broad set of training images, optimizing a robust convolutional neural network that is agnostic to MRI contrasts and acquisition schemes.

\section*{Acknowledgments}

The authors thank Douglas Greve, Lilla Z{\"o}llei, and David Salat for sharing FSM, infant, and ASL data, respectively, and for providing a mechanism for distributing our reference dataset. Support for this research was provided in part by a BRAIN Initiative Cell Census Network grant [U01 MH117023], the National Institute of Biomedical Imaging and Bioengineering [P41 EB015896, R01 EB023281, R21 EB018907, R01 EB019956, P41 EB030006], the National Institute of Child Health and Human Development [K99 HD101553], the National Institute on Aging [R56 AG064027, R01 AG016495, R01 AG070988], the National Institute of Mental Health [RF1 MH121885, RF1 MH123195], the National Institute of Neurological Disorders and Stroke [R01 NS070963, R01 NS083534, R01 NS105820], and was made possible by the resources provided by Shared Instrumentation Grants [S10 RR023401, S10 RR019307, S10 RR023043]. Additional support was provided by the NIH Blueprint for Neuroscience Research [U01 MH093765], part of the multi-institutional Human Connectome Project.
Furthermore, this research project benefited from computational hardware generously provided by the Massachusetts Life Sciences Center (\url{https://www.masslifesciences.com}).
Lastly, Bruce Fischl has a financial interest in CorticoMetrics, a company whose medical pursuits focus on brain imaging and measurement technologies. This interest is reviewed and managed by Massachusetts General Hospital and Mass General Brigham in accordance with their conflict of interest policies.

{
\bibliographystyle{plainnat}
\footnotesize
\bibliography{ref}
}

\newpage
\pagenumbering{gobble}

\setcounter{table}{0}
\renewcommand{\thetable}{S\arabic{table}}
\setcounter{figure}{0}
\renewcommand{\thefigure}{S\arabic{figure}}

\begin{table*}[]
\centering
\small
\begin{tabular}{lllllllll}

\multicolumn{8}{c}{\textbf{Max Surface Distance (mm)}} \\
[1ex] \toprule \\ [-2ex]
                 &   SynthStrip  &  ROBEX  &  BET  &  3DSS  &  BEaST  &  FSW  &  DMBE  \\
[0.5ex] \toprule \\ [-2.5ex]
IXI T1w          &  $13.5 \pm 7.0$  &  $17.4 \pm 6.8$  &  $17.2 \pm 11.8$  &  $29.3 \pm 13.2$  &  $20.6 \pm 6.1$  &  $20.7 \pm 17.1$  &  $39.5 \pm 27.1$  \\
&    &  {\footnotesize $3.9 \times 10^{-11}$}  &  {\footnotesize $2.0 \times 10^{-2}$}  &  {\footnotesize $1.2 \times 10^{-9}$}  &  {\footnotesize $5.7 \times 10^{-22}$}  &  {\footnotesize $8.1 \times 10^{-4}$}  &  {\footnotesize $1.8 \times 10^{-8}$}  \\
\hline \\ [-2.3ex]
FSM T1w          &  $12.3 \pm 2.7$  &  $16.0 \pm 2.2$  &  $85.4 \pm 22.5$  &  $31.0 \pm 14.4$  &  $19.3 \pm 2.3$  &  $22.4 \pm 15.6$  &  $70.7 \pm 26.3$  \\
&    &  {\footnotesize $6.1 \times 10^{-9}$}  &  {\footnotesize $1.9 \times 10^{-20}$}  &  {\footnotesize $2.7 \times 10^{-9}$}  &  {\footnotesize $2.9 \times 10^{-16}$}  &  {\footnotesize $3.4 \times 10^{-4}$}  &  {\footnotesize $9.1 \times 10^{-16}$}  \\
\hline \\ [-2.3ex]
ASL T1w          &  $12.2 \pm 2.7$  &  $15.0 \pm 3.6$  &  $16.1 \pm 4.4$  &  $32.6 \pm 14.7$  &  $20.9 \pm 5.4$  &  $11.7 \pm 1.4$  &  $52.6 \pm 20.9$  \\
&    &  {\footnotesize $2.5 \times 10^{-4}$}  &  {\footnotesize $1.8 \times 10^{-5}$}  &  {\footnotesize $7.0 \times 10^{-11}$}  &  {\footnotesize $1.3 \times 10^{-12}$}  &  {\footnotesize {\color{orange} $2.2 \times 10^{-1}$}}  &  {\footnotesize $5.4 \times 10^{-16}$}  \\
\hline \\ [-2.3ex]
QIN T1w          &  $16.9 \pm 5.3$  &  $21.9 \pm 6.0$  &  $22.7 \pm 7.2$  &  $39.3 \pm 12.4$  &  $29.0 \pm 19.1$  &  $42.1 \pm 20.1$  &  $32.8 \pm 10.0$  \\
&    &  {\footnotesize $3.0 \times 10^{-5}$}  &  {\footnotesize $5.6 \times 10^{-7}$}  &  {\footnotesize $1.5 \times 10^{-17}$}  &  {\footnotesize $2.3 \times 10^{-5}$}  &  {\footnotesize $1.8 \times 10^{-12}$}  &  {\footnotesize $1.3 \times 10^{-15}$}  \\
\hline \\ [-2.3ex]
IXI T2w          &  $13.1 \pm 4.0$  &  $16.9 \pm 4.8$  &  $22.6 \pm 7.7$  &  $31.5 \pm 13.6$  &  $66.2 \pm 21.5$  &  -  &  $122.4 \pm 17.7$  \\
&    &  {\footnotesize $1.2 \times 10^{-6}$}  &  {\footnotesize $6.9 \times 10^{-11}$}  &  {\footnotesize $1.8 \times 10^{-12}$}  &  {\footnotesize $5.9 \times 10^{-23}$}  &  -  &  {\footnotesize $3.6 \times 10^{-39}$}  \\
\hline \\ [-2.3ex]
FSM T2w          &  $11.2 \pm 2.6$  &  $13.7 \pm 4.8$  &  $21.4 \pm 8.9$  &  $22.9 \pm 9.9$  &  $59.2 \pm 29.2$  &  -  &  $136.9 \pm 21.8$  \\
&    &  {\footnotesize $9.9 \times 10^{-3}$}  &  {\footnotesize $1.8 \times 10^{-7}$}  &  {\footnotesize $4.5 \times 10^{-8}$}  &  {\footnotesize $2.0 \times 10^{-11}$}  &  -  &  {\footnotesize $1.1 \times 10^{-28}$}  \\
\hline \\ [-2.3ex]
QIN T2w          &  $23.2 \pm 9.8$  &  $29.1 \pm 8.6$  &  $28.6 \pm 9.7$  &  $51.3 \pm 23.1$  &  $57.9 \pm 20.4$  &  -  &  $67.5 \pm 19.7$  \\
&    &  {\footnotesize $5.8 \times 10^{-5}$}  &  {\footnotesize $2.2 \times 10^{-5}$}  &  {\footnotesize $2.3 \times 10^{-8}$}  &  {\footnotesize $1.1 \times 10^{-11}$}  &  -  &  {\footnotesize $2.3 \times 10^{-16}$}  \\
\hline \\ [-2.3ex]
QIN FLAIR        &  $15.0 \pm 2.4$  &  $20.8 \pm 5.6$  &  $18.3 \pm 6.7$  &  $37.3 \pm 12.1$  &  $26.3 \pm 8.5$  &  $42.6 \pm 10.3$  &  $39.9 \pm 8.2$  \\
&    &  {\footnotesize $1.7 \times 10^{-3}$}  &  {\footnotesize {\color{orange} $6.0 \times 10^{-2}$}}  &  {\footnotesize $2.4 \times 10^{-6}$}  &  {\footnotesize $8.7 \times 10^{-5}$}  &  {\footnotesize $2.5 \times 10^{-8}$}  &  {\footnotesize $1.4 \times 10^{-9}$}  \\
\hline \\ [-2.3ex]
IXI PDw          &  $13.2 \pm 4.1$  &  $13.7 \pm 4.2$  &  $18.5 \pm 6.0$  &  $34.5 \pm 16.1$  &  $52.8 \pm 24.4$  &  $43.2 \pm 21.1$  &  $42.9 \pm 12.9$  \\
&    &  {\footnotesize {\color{orange} $3.1 \times 10^{-1}$}}  &  {\footnotesize $2.6 \times 10^{-7}$}  &  {\footnotesize $1.3 \times 10^{-12}$}  &  {\footnotesize $4.4 \times 10^{-15}$}  &  {\footnotesize $1.3 \times 10^{-12}$}  &  {\footnotesize $1.9 \times 10^{-19}$}  \\
\hline \\ [-2.3ex]
FSM PDw          &  $10.9 \pm 1.7$  &  $10.3 \pm 3.7$  &  $14.8 \pm 15.4$  &  $15.6 \pm 9.2$  &  $26.9 \pm 10.4$  &  $98.3 \pm 18.0$  &  $94.4 \pm 6.0$  \\
&    &  {\footnotesize {\color{orange} $3.9 \times 10^{-1}$}}  &  {\footnotesize {\color{orange} $1.6 \times 10^{-1}$}}  &  {\footnotesize $9.1 \times 10^{-3}$}  &  {\footnotesize $1.4 \times 10^{-9}$}  &  {\footnotesize $7.7 \times 10^{-23}$}  &  {\footnotesize $1.6 \times 10^{-36}$}  \\
\hline \\ [-2.3ex]
IXI MRA          &  $39.3 \pm 13.6$  &  $42.2 \pm 13.7$  &  $53.9 \pm 11.4$  &  $39.2 \pm 15.3$  &  $36.8 \pm 12.3$  &  -  &  $92.4 \pm 28.0$  \\
&    &  {\footnotesize $4.6 \times 10^{-3}$}  &  {\footnotesize $2.2 \times 10^{-6}$}  &  {\footnotesize {\color{orange} $9.8 \times 10^{-1}$}}  &  {\footnotesize {\color{orange} $4.2 \times 10^{-1}$}}  &  -  &  {\footnotesize $4.1 \times 10^{-16}$}  \\
\hline \\ [-2.3ex]
FSM qT1          &  $11.6 \pm 3.1$  &  $61.0 \pm 25.0$  &  $132.3 \pm 19.0$  &  $89.1 \pm 6.0$  &  $84.2 \pm 33.5$  &  $91.5 \pm 33.6$  &  $130.1 \pm 27.0$  \\
&    &  {\footnotesize $7.8 \times 10^{-12}$}  &  {\footnotesize $2.6 \times 10^{-26}$}  &  {\footnotesize $5.0 \times 10^{-33}$}  &  {\footnotesize $8.3 \times 10^{-13}$}  &  {\footnotesize $9.5 \times 10^{-14}$}  &  {\footnotesize $1.0 \times 10^{-21}$}  \\
\hline \\ [-2.3ex]
ASL EPI          &  $17.3 \pm 6.5$  &  $25.3 \pm 4.4$  &  $18.9 \pm 5.8$  &  $32.7 \pm 8.8$  &  $55.3 \pm 23.3$  &  $52.9 \pm 11.5$  &  $23.8 \pm 6.9$  \\
&    &  {\footnotesize $3.5 \times 10^{-7}$}  &  {\footnotesize {\color{orange} $1.7 \times 10^{-1}$}}  &  {\footnotesize $8.2 \times 10^{-13}$}  &  {\footnotesize $2.9 \times 10^{-12}$}  &  {\footnotesize $6.3 \times 10^{-19}$}  &  {\footnotesize $3.7 \times 10^{-7}$}  \\
\hline \\ [-2.3ex]
Infant T1w       &  $38.7 \pm 22.2$  &  $44.1 \pm 20.0$  &  $64.1 \pm 19.4$  &  $78.8 \pm 26.7$  &  $68.6 \pm 40.8$  &  $65.1 \pm 27.6$  &  $60.5 \pm 16.4$  \\
&    &  {\footnotesize {\color{orange} $9.9 \times 10^{-2}$}}  &  {\footnotesize $6.4 \times 10^{-3}$}  &  {\footnotesize $1.7 \times 10^{-4}$}  &  {\footnotesize $1.0 \times 10^{-2}$}  &  {\footnotesize $7.5 \times 10^{-3}$}  &  {\footnotesize $2.6 \times 10^{-3}$}  \\
\hline \\ [-2.3ex]
IXI DWI          &  $17.3 \pm 5.2$  &  $23.0 \pm 5.3$  &  $21.5 \pm 3.5$  &  $33.5 \pm 6.7$  &  $45.1 \pm 24.6$  &  $37.9 \pm 9.0$  &  $36.4 \pm 7.5$  \\
&    &  {\footnotesize $1.6 \times 10^{-5}$}  &  {\footnotesize $1.4 \times 10^{-4}$}  &  {\footnotesize $8.1 \times 10^{-13}$}  &  {\footnotesize $1.1 \times 10^{-6}$}  &  {\footnotesize $1.2 \times 10^{-11}$}  &  {\footnotesize $2.2 \times 10^{-16}$}  \\
\hline \\ [-2.3ex]
CIM PET          &  $12.0 \pm 2.0$  &  $15.7 \pm 5.2$  &  $49.9 \pm 17.2$  &  $20.2 \pm 12.5$  &  $147.3 \pm 25.3$  &  $51.8 \pm 9.6$  &  $72.6 \pm 16.0$  \\
&    &  {\footnotesize $4.7 \times 10^{-3}$}  &  {\footnotesize $2.6 \times 10^{-8}$}  &  {\footnotesize $1.2 \times 10^{-2}$}  &  {\footnotesize $7.2 \times 10^{-14}$}  &  {\footnotesize $5.4 \times 10^{-12}$}  &  {\footnotesize $8.1 \times 10^{-13}$}  \\
\hline \\ [-2.3ex]
CIM CT           &  $12.3 \pm 2.5$  &  $24.7 \pm 4.7$  &  $123.9 \pm 13.7$  &  $58.0 \pm 7.2$  &  $155.7 \pm 22.3$  &  $105.1 \pm 20.7$  &  $132.9 \pm 24.0$  \\
&    &  {\footnotesize $5.9 \times 10^{-10}$}  &  {\footnotesize $1.0 \times 10^{-18}$}  &  {\footnotesize $7.1 \times 10^{-16}$}  &  {\footnotesize $3.0 \times 10^{-15}$}  &  {\footnotesize $2.6 \times 10^{-12}$}  &  {\footnotesize $1.5 \times 10^{-14}$}  \\
\bottomrule \\ [0ex]

\end{tabular}
\caption{Skull-stripping accuracy across datasets, as measured by the mean Hausdorff distance ($\pm$~SD) between computed and ground-truth binary brain masks.~\textit{p}-values comparing baseline with SythStrip results are presented below each score. Across each dataset, SynthStrip significantly outperforms most baselines except those with~\textit{p}-values in orange (\textit{p}~$>$ 0.05).}
\label{table_dist_max}
\end{table*}

\begin{table*}[]
\centering
\small
\begin{tabular}{lllllllll}

\multicolumn{8}{c}{\textbf{Volume Difference (\%)}} \\
[1ex] \toprule \\ [-2ex]
                 &   SynthStrip  &  ROBEX  &  BET  &  3DSS  &  BEaST  &  FSW  &  DMBE  \\
[0.5ex] \toprule \\ [-2.5ex]
IXI T1w          &  $2.8 \pm 1.9$  &  $5.8 \pm 1.9$  &  $5.2 \pm 6.0$  &  $7.0 \pm 3.2$  &  $11.9 \pm 2.2$  &  $16.0 \pm 11.7$  &  $6.9 \pm 3.6$  \\
&    &  {\footnotesize $2.0 \times 10^{-11}$}  &  {\footnotesize $1.1 \times 10^{-2}$}  &  {\footnotesize $5.9 \times 10^{-10}$}  &  {\footnotesize $3.2 \times 10^{-26}$}  &  {\footnotesize $1.4 \times 10^{-9}$}  &  {\footnotesize $1.1 \times 10^{-9}$}  \\
\hline \\ [-2.3ex]
FSM T1w          &  $1.1 \pm 1.0$  &  $7.5 \pm 1.5$  &  $104.7 \pm 41.7$  &  $5.7 \pm 4.4$  &  $14.5 \pm 1.2$  &  $12.5 \pm 10.0$  &  $7.5 \pm 4.1$  \\
&    &  {\footnotesize $6.7 \times 10^{-21}$}  &  {\footnotesize $2.2 \times 10^{-17}$}  &  {\footnotesize $4.1 \times 10^{-7}$}  &  {\footnotesize $1.4 \times 10^{-35}$}  &  {\footnotesize $3.3 \times 10^{-8}$}  &  {\footnotesize $2.3 \times 10^{-11}$}  \\
\hline \\ [-2.3ex]
ASL T1w          &  $3.2 \pm 1.6$  &  $4.1 \pm 1.5$  &  $8.6 \pm 2.4$  &  $9.0 \pm 4.8$  &  $19.4 \pm 7.0$  &  $8.5 \pm 2.4$  &  $8.3 \pm 4.0$  \\
&    &  {\footnotesize $6.1 \times 10^{-3}$}  &  {\footnotesize $5.4 \times 10^{-18}$}  &  {\footnotesize $5.3 \times 10^{-9}$}  &  {\footnotesize $1.9 \times 10^{-17}$}  &  {\footnotesize $6.3 \times 10^{-12}$}  &  {\footnotesize $7.7 \times 10^{-9}$}  \\
\hline \\ [-2.3ex]
QIN T1w          &  $1.9 \pm 1.5$  &  $8.2 \pm 13.1$  &  $6.4 \pm 5.2$  &  $8.0 \pm 7.3$  &  $17.6 \pm 22.2$  &  $83.2 \pm 185.3$  &  $15.5 \pm 11.6$  \\
&    &  {\footnotesize $6.8 \times 10^{-4}$}  &  {\footnotesize $4.0 \times 10^{-8}$}  &  {\footnotesize $6.8 \times 10^{-8}$}  &  {\footnotesize $4.3 \times 10^{-6}$}  &  {\footnotesize $2.3 \times 10^{-3}$}  &  {\footnotesize $1.5 \times 10^{-11}$}  \\
\hline \\ [-2.3ex]
IXI T2w          &  $2.6 \pm 1.3$  &  $17.0 \pm 6.8$  &  $8.7 \pm 10.4$  &  $3.8 \pm 3.2$  &  $80.4 \pm 143.0$  &  -  &  $96.2 \pm 5.0$  \\
&    &  {\footnotesize $2.6 \times 10^{-19}$}  &  {\footnotesize $2.7 \times 10^{-4}$}  &  {\footnotesize $2.5 \times 10^{-2}$}  &  {\footnotesize $3.9 \times 10^{-4}$}  &  {\footnotesize -}  &  {\footnotesize $8.9 \times 10^{-64}$}  \\
\hline \\ [-2.3ex]
FSM T2w          &  $1.1 \pm 0.9$  &  $11.9 \pm 4.2$  &  $5.7 \pm 4.0$  &  $4.0 \pm 3.2$  &  $70.2 \pm 95.7$  &  -  &  $96.0 \pm 5.3$  \\
&    &  {\footnotesize $2.2 \times 10^{-16}$}  &  {\footnotesize $1.6 \times 10^{-7}$}  &  {\footnotesize $1.3 \times 10^{-5}$}  &  {\footnotesize $1.4 \times 10^{-4}$}  &  {\footnotesize -}  &  {\footnotesize $1.7 \times 10^{-45}$}  \\
\hline \\ [-2.3ex]
QIN T2w          &  $3.0 \pm 2.9$  &  $73.3 \pm 216.1$  &  $57.7 \pm 219.1$  &  $172.9 \pm 353.4$  &  $91.5 \pm 213.6$  &  -  &  $31.2 \pm 22.4$  \\
&    &  {\footnotesize {\color{orange} $5.2 \times 10^{-2}$}}  &  {\footnotesize {\color{orange} $1.3 \times 10^{-1}$}}  &  {\footnotesize $5.2 \times 10^{-3}$}  &  {\footnotesize $1.5 \times 10^{-2}$}  &  {\footnotesize -}  &  {\footnotesize $1.7 \times 10^{-9}$}  \\
\hline \\ [-2.3ex]
QIN FLAIR        &  $1.8 \pm 1.0$  &  $9.6 \pm 3.8$  &  $2.3 \pm 2.3$  &  $4.8 \pm 2.0$  &  $17.0 \pm 7.3$  &  $34.6 \pm 19.6$  &  $13.0 \pm 4.3$  \\
&    &  {\footnotesize $1.7 \times 10^{-6}$}  &  {\footnotesize {\color{orange} $4.0 \times 10^{-1}$}}  &  {\footnotesize $1.7 \times 10^{-5}$}  &  {\footnotesize $3.5 \times 10^{-7}$}  &  {\footnotesize $5.9 \times 10^{-6}$}  &  {\footnotesize $2.3 \times 10^{-8}$}  \\
\hline \\ [-2.3ex]
IXI PDw          &  $2.3 \pm 1.1$  &  $7.2 \pm 3.3$  &  $4.4 \pm 3.1$  &  $3.6 \pm 3.0$  &  $52.0 \pm 83.3$  &  $105.0 \pm 229.4$  &  $8.9 \pm 5.7$  \\
&    &  {\footnotesize $9.2 \times 10^{-14}$}  &  {\footnotesize $1.2 \times 10^{-5}$}  &  {\footnotesize $1.1 \times 10^{-2}$}  &  {\footnotesize $1.2 \times 10^{-4}$}  &  {\footnotesize $2.9 \times 10^{-3}$}  &  {\footnotesize $1.2 \times 10^{-10}$}  \\
\hline \\ [-2.3ex]
FSM PDw          &  $3.0 \pm 1.8$  &  $4.7 \pm 3.3$  &  $7.5 \pm 22.7$  &  $4.8 \pm 2.4$  &  $15.5 \pm 13.6$  &  $102.3 \pm 45.2$  &  $37.4 \pm 14.1$  \\
&    &  {\footnotesize $4.4 \times 10^{-2}$}  &  {\footnotesize {\color{orange} $2.8 \times 10^{-1}$}}  &  {\footnotesize $4.0 \times 10^{-5}$}  &  {\footnotesize $1.2 \times 10^{-5}$}  &  {\footnotesize $3.6 \times 10^{-13}$}  &  {\footnotesize $1.9 \times 10^{-14}$}  \\
\hline \\ [-2.3ex]
IXI MRA          &  $1.7 \pm 1.1$  &  $44.4 \pm 12.1$  &  $52.7 \pm 16.6$  &  $8.6 \pm 2.1$  &  $111.6 \pm 296.2$  &  -  &  $88.4 \pm 15.1$  \\
&    &  {\footnotesize $7.2 \times 10^{-29}$}  &  {\footnotesize $1.5 \times 10^{-26}$}  &  {\footnotesize $3.6 \times 10^{-29}$}  &  {\footnotesize $1.2 \times 10^{-2}$}  &  {\footnotesize -}  &  {\footnotesize $5.3 \times 10^{-39}$}  \\
\hline \\ [-2.3ex]
FSM qT1         &  $1.1 \pm 1.0$  &  $173.8 \pm 191.0$  &  $243.2 \pm 116.4$  &  $262.2 \pm 139.2$  &  $98.8 \pm 166.6$  &  $153.8 \pm 193.5$  &  $76.6 \pm 15.2$  \\
&    &  {\footnotesize $2.0 \times 10^{-5}$}  &  {\footnotesize $8.5 \times 10^{-13}$}  &  {\footnotesize $1.1 \times 10^{-11}$}  &  {\footnotesize $2.7 \times 10^{-3}$}  &  {\footnotesize $1.2 \times 10^{-4}$}  &  {\footnotesize $2.0 \times 10^{-23}$}  \\
\hline \\ [-2.3ex]
ASL EPI          &  $5.8 \pm 2.3$  &  $24.4 \pm 14.1$  &  $7.2 \pm 4.3$  &  $3.9 \pm 2.8$  &  $77.6 \pm 93.1$  &  $81.8 \pm 27.4$  &  $6.5 \pm 3.4$  \\
&    &  {\footnotesize $4.6 \times 10^{-10}$}  &  {\footnotesize {\color{orange} $1.5 \times 10^{-1}$}}  &  {\footnotesize {\color{orange} $8.3 \times 10^{-2}$}}  &  {\footnotesize $1.4 \times 10^{-5}$}  &  {\footnotesize $1.6 \times 10^{-20}$}  &  {\footnotesize {\color{orange} $1.8 \times 10^{-1}$}}  \\
\hline \\ [-2.3ex]
Infant T1w       &  $7.4 \pm 3.5$  &  $34.7 \pm 65.7$  &  $142.4 \pm 154.6$  &  $106.2 \pm 124.6$  &  $228.0 \pm 358.7$  &  $259.4 \pm 294.2$  &  $19.8 \pm 11.8$  \\
&    &  {\footnotesize {\color{orange} $1.2 \times 10^{-1}$}}  &  {\footnotesize $3.7 \times 10^{-3}$}  &  {\footnotesize $7.0 \times 10^{-3}$}  &  {\footnotesize $3.0 \times 10^{-2}$}  &  {\footnotesize $4.6 \times 10^{-3}$}  &  {\footnotesize $3.7 \times 10^{-4}$}  \\
\hline \\ [-2.3ex]
IXI DWI          &  $5.5 \pm 3.5$  &  $32.0 \pm 18.0$  &  $10.8 \pm 4.5$  &  $17.0 \pm 3.9$  &  $37.8 \pm 33.0$  &  $84.2 \pm 167.9$  &  $31.9 \pm 4.2$  \\
&    &  {\footnotesize $1.2 \times 10^{-8}$}  &  {\footnotesize $5.4 \times 10^{-12}$}  &  {\footnotesize $1.0 \times 10^{-18}$}  &  {\footnotesize $5.6 \times 10^{-6}$}  &  {\footnotesize $1.3 \times 10^{-2}$}  &  {\footnotesize $2.1 \times 10^{-31}$}  \\
\hline \\ [-2.3ex]
CIM PET          &  $3.6 \pm 3.6$  &  $23.3 \pm 14.1$  &  $58.1 \pm 29.1$  &  $8.3 \pm 20.1$  &  $174.9 \pm 275.5$  &  $120.3 \pm 207.9$  &  $29.2 \pm 20.9$  \\
&    &  {\footnotesize $2.0 \times 10^{-6}$}  &  {\footnotesize $7.4 \times 10^{-8}$}  &  {\footnotesize {\color{orange} $2.7 \times 10^{-1}$}}  &  {\footnotesize $1.4 \times 10^{-2}$}  &  {\footnotesize $2.4 \times 10^{-2}$}  &  {\footnotesize $5.4 \times 10^{-5}$}  \\
\hline \\ [-2.3ex]
CIM CT           &  $11.5 \pm 2.0$  &  $70.5 \pm 7.7$  &  $243.9 \pm 40.4$  &  $132.0 \pm 16.2$  &  $157.4 \pm 288.4$  &  $313.6 \pm 292.4$  &  $64.3 \pm 24.5$  \\
&    &  {\footnotesize $6.3 \times 10^{-20}$}  &  {\footnotesize $4.7 \times 10^{-16}$}  &  {\footnotesize $2.6 \times 10^{-18}$}  &  {\footnotesize $4.0 \times 10^{-2}$}  &  {\footnotesize $2.4 \times 10^{-4}$}  &  {\footnotesize $1.2 \times 10^{-8}$}  \\
\bottomrule \\ [0ex]

\end{tabular}
\caption{Skull-stripping accuracy across datasets, as measured by the mean percent difference ($\pm$~SD) in volume between computed and ground-truth binary brain masks.~\textit{p}-values comparing baseline with SythStrip results are presented below each score. Across each dataset, SynthStrip significantly outperforms most baselines except those with~\textit{p}-values in orange, for which~\textit{p}~$>$ 0.05.}
\label{table_voldiff}
\end{table*}

\begin{figure*}[]
    \centering
    \includegraphics[width=\textwidth]{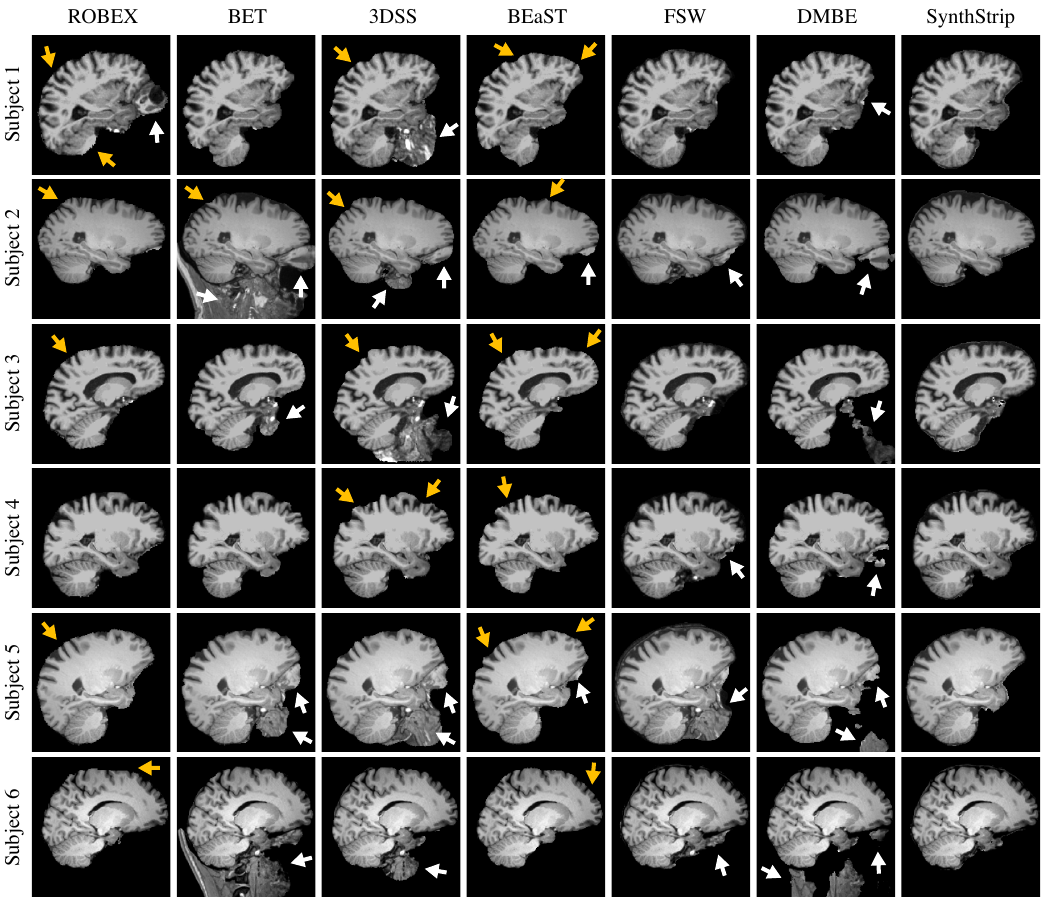}
    \caption{Comparison of representative skull-stripping errors for each method across six individual test scans. White arrows indicate over-labeling of the brain mask, while orange arrows indicate removal of brain matter.}
    \label{fig:failure_grid}
\end{figure*}

\end{document}